\documentclass[a4paper,12pt]{article}

\usepackage{amssymb,amsmath,mathbbol,mathrsfs}
\usepackage[usenames,dvipsnames]{color}
\usepackage{hyperref}
\usepackage{xcolor}
\usepackage{stmaryrd}
\usepackage{authblk}
\usepackage{framed}
\usepackage{empheq}
\usepackage{slashed}
\usepackage{hyphenat}
\usepackage{cite}

\usepackage{marginnote}    

\usepackage[left=.8in,right=.8in,top=.8in,bottom=.8in]{geometry}                  

\usepackage{sectsty}
\allsectionsfont{\sffamily\mdseries\upshape} 
\usepackage{tocloft}

\makeatletter
\renewenvironment{abstract}{%
    \if@twocolumn
      \section*{\abstractname}%
    \else 
      \begin{center}%
        {\bfseries\sffamily\abstractname\vspace{\z@}}
      \end{center}%
      \quotation
    \fi}
    {\if@twocolumn\else\endquotation\fi}
\makeatother

\numberwithin{equation}{section}
5

\hypersetup{
	colorlinks=true,         
	linkcolor=MidnightBlue,          
	citecolor=BrickRed,        
	urlcolor=MidnightBlue            
}

\newcommand{\be}{\begin{equation}}
\newcommand{\ee}{\end{equation}}

\newcommand{\fg}{{\mathfrak{g}}}

\newcommand{\fLie}{{\mathbb{L}}}
\newcommand{\fI}{{\mathbb{i}}}
\newcommand{\F}{{\Phi}}
\renewcommand{\d}{{\mathrm{d}}}
\newcommand{\D}{{\mathrm{D}}}

\newcommand{\SU}{{\mathrm{SU}}}
\newcommand{\YM}{{\text{YM}}}
\newcommand{\G}{{\mathcal{G}}}

\newcommand{\pp}{{\partial}}
\newcommand{\fvf}{{\mathbb{X}}}
\newcommand{\lvf}{{\xi}}

\newcommand{\FYM}{{\Phi_\text{YM}}}

\newcommand{\PPexp}{{\mathbb{P}\hspace{-1pt}\exp}}
\newcommand{\fG}{{\mathrm{Lie}(\G)}}
\newcommand{\fX}{{\mathfrak{X}}}

\renewcommand{\bar}{\overline}
\newcommand{\dd}{{\mathbb{d}}}

\renewcommand{\hat}{\widehat}

\newcommand{\cint}{{\int\kern-.87em{<}}}
\newcommand{\sint}{{\int\kern-.75em{\sim}}}
\newcommand{\fint}{{\int\kern-1.00em{\int}}}

\newcommand{\bb}{\mathbb}

\newcommand{\tr}{\text{tr}}

\renewcommand{\#}{\sharp}

\let\oldmarginpar\marginpar
\renewcommand\marginpar[1]{\oldmarginpar{\color{red}\raggedright\footnotesize #1}}

\title{\sffamily Gauging the boundary in field-space}
\author[a]{ \sffamily  Henrique Gomes\thanks{gomes.ha@gmail.com, hdag2@cam.ac.uk}}
\affil[a]{Trinity College\\ Cambridge, CB2 1TQ, United Kingdom}

\begin{document}
\maketitle

\abstract{ Local gauge theories are in a complicated relationship with boundaries. Whereas fixing the gauge can often shave off unwanted redundancies,  the coupling of different  bounded regions requires the use of gauge-variant elements. Therefore, coupling is inimical to gauge-fixing, as usually understood. This resistance to gauge-fixing has led some to declare the coupling of subsystems to be the \textit{raison d'\^etre} of gauge \cite{RovelliGauge2013}.  

Indeed, while gauge-fixing is entirely unproblematic for a single region without boundary, it introduces arbitrary boundary conditions on the gauge degrees of freedom themselves---these  conditions  lack a physical interpretation when they are not functionals of the original fields. 
 Such arbitrary boundary choices creep into the calculation of charges through Noether's second theorem, muddling the assignment of physical charges to local gauge symmetries. The confusion brewn by gauge at boundaries is well-known, and must be contended with both conceptually and technically. 

  It may seem natural to replace the arbitrary boundary choice with new degrees of freedom, for using such a device we resolve some of these confusions while 
   leaving no gauge-dependence on the computation of Noether charges \cite{DonnellyFreidel}. But, concretely, such boundary degrees of freedom are rather arbitrary---they have no relation to the original field-content of the field theory. How should we conceive of them? 
 
Here I will explicate the problems mentioned above and illustrate a possible resolution:  in a recent series of papers \cite{GomesRiello2016, GomesRiello2018, GomesHopfRiello} the notion of a connection-form was put forward and  implemented in the field-space of gauge theories. Using this tool, a modified version of symplectic geometry---here called `horizontal'---is possible.  Independently of boundary conditions, this formalism bestows to each region a physically salient, relational notion of charge: the horizontal Noether charge. Meanwhile, as required, the connection-form mediates a composition of regions, one compatible with the attribution of horizontal Noether charges to each region. 
The aims of this paper are to highlight the boundary issues in the treatment of gauge, and to discuss how gauge theory may be conceptually clarified in light of a resolution to these issues.  }

\newpage
{\hypersetup{	linkcolor=black, }\tableofcontents}

\begin{center}
	\rule{8cm}{0.4pt}
\end{center}


 \textbf {Roadmap:} In section \ref{sec:intro}, I criticize the usual construal of gauge-invariance as descriptive redundancy, by buiding on Rovelli's view that gauge variables are needed to describe the relations between subsystems, and thus, in a field-theory, to  describe boundaries.
 Which gauge variables are to be kept, and how they couple is not made explicit by Rovelli. I identify the very flexible notion of a `perspective' as the right tool to meet Rovelli's requirements. The challenge then becomes one of keeping the perspectives while still being able to distinguish physical processes from pure gauge ones. 
 
 In section \ref{sec:hor_symp_geom}, this challenge is met. I elaborate on Rovelli's intuition, by providing specific, regional, gauge-covariant `handles', used to couple different regions. These handles are given by the \textit{relational connection-form}. { If the standard finite-dimensional connection-form arises by augmenting spacetime derivatives to be covariant under spacetime-dependent gauge transformations,  the relational connection-form arises from augmenting field-variations to be covariant under field-dependent gauge transformations \cite{GomesRiello2016}. The point is that, at boundaries, the distinction between physical and gauge degrees of freedom is muddied, and therefore gauge-transformations there must adopt  field-dependence. }
 
 Alternatively, the guiding requirement for the construction of the relational connection-form  is  a harmonious melding of  regional and global perspectives. Most importantly, retaining gauge-variant elements \textit{in this manner} is compatible with the attribution  of a (relational) physical status to processes, and this is true both globally and within bounded regions, and with or  without imposing boundary conditions.  Lastly, I show that the ensuing  notions of regional relationalism are different from other attempts at resolving  the problem posed by gauge symmetries for bounded regions. The distinguishing criterion is what I consider to be the `acid test' of local gauge theories in bounded regions: does the theory license only those regional charges which depend solely on the original field content? In a satisfactory theory, the regional charges should only depend on the original field content. In section \ref{sec:connection_form},  I introduce explicit examples of relational connection-forms, and show that the ensuing horizontal symplectic geometry  passes this `acid test'.

\section{Gauge and boundary: a complicated relationship.}\label{sec:intro}

 In section \ref{sec:Rovelli} I argue that gauge is a matter of alternative descriptions, and that  keeping these different alternatives in play is indispensable  when we recognize that the system described is not the entire cosmos. These different descriptions are the `perspectives'.
 
  In section \ref{sec:contraRovelli} I argue that we need a more precise definition than Rovelli provides of what types of ``gauge-variant handles''  are needed to couple regions. Harmonizing regional and global perspectives could do the trick. 
  
  In section \ref{sec:gf}, using gauge-fixings, I show how in the field-theory context boundaries retain the difficulties about coupling identified by Rovelli. I show that the underlying reason is that boundaries drive a wedge between the notion of gauge as: i) a matter of alternative descriptions of a system and as ii) a matter of applications of an abstract group of gauge transformations. It is this wedge that makes room for the---erroneous, in my view---introduction of extra physical degrees of freedom at boundaries, an issue tackled in  \ref{sec:hist_intro}. In that subsection,  we take stock of our argument so far,  then move on to the consequences of this argument for Noether charges, and Donnelly and Freidel's proposal of physical edge-modes \cite{DonnellyFreidel}.

\subsection{What is gauge?}\label{sec:Rovelli}
\subsubsection{\textit{On perspectives} --}

{ 

\begin{quote}``Like moths attracted to a bright light, philosophers are drawn to glitz. So in discussing
the notions of ‘gauge’, ‘gauge freedom’, and ‘gauge theories’, they have
tended to focus on examples such as Yang–Mills theories and on the mathematical
apparatus of fibre bundles. But while Yang–Mills theories are crucial to modern
elementary particle physics, they are only a special case of a much broader class of
gauge theories. And while the fibre bundle apparatus turned out, in retrospect, to be
the right formalism to illuminate the structure of Yang–Mills theories, the strength
of this apparatus is also its weakness: the fibre bundle formalism is very flexible
and general, and, as such, fibre bundles can be seen lurking under, over, and around
every bush.'' -- J. Earman \cite{Earman_ode}\end{quote}
I am  one of those who is attracted to the ``glitz'' of fiber bundles. However, I believe there is more to the glitz than glamorous display; the generality  of fiber bundles  captures a defining feature of gauge theories. In simple terms, fiber bundles tell us  }
  a `gauge system' is one that allows different descriptions of itself---different perspectives.  Relating  these perspectives is then the job of gauge transformations. This broad characterization doesn't make reference to action functionals or equations of motion; they come in later in the  definition of the  dynamics of the system.

  The  isospin fields---which orginally motivated Yang and Mills to introduce the local non-Abelian gauge groups \cite{YangMills}---is a good example; admitting descriptions under different labeling of particles with isotopic spin.\footnote{ But so is the lamp on my desk, which can be visually described by several different  perspectives. See chapter 3, pages 66-75, of Van Fraassen \cite{VanFraasen} for a philosophical discussion connecting visual perspectives, Cartesian frames of reference and Lorentz-invariant descriptions (all global transformations, as opposed to the Yang-Mills example). Such analogies can lend at least a smidgeon of plausibility to this broad painting of gauge, even if  now gauge systems may ``lurk under every bush''. Admittedly,  a strict adherence to this characterization of gauge systems might overextend its usefulness by including too many types of systems under the umbrella of `gauge'. Rest assured,  we will stay dry; I won't stray  from standard examples.}

   Underlying this simple definition lies a subtle distinction between the gauge (or symmetry) group, and the  perspectives. The perspective is the description of the state from a given viewpoint. The perspectives emphasize how the same system is viewed within different frames, whereas the symmetry groups emphasize the abstract `rotations' between frames, or viewpoints.  A `frame' would be the characterization of the coordinates themselves, whereas the perspective characterizes  the system as described in that frame. For instance, a frame could be a choice of basis of some vector space, whereas the perspective is given by the components of the system under consideration, in that basis.  As we will see,  a change of frames is identified with a group element and generically corresponds to a change in perspectives.
   
Although changes of perspective could be said to be more concrete than the abstract definition of gauge symmetries,  there are more important, objective, albeit subtle, differences  between `frame' and `perspective'. 
Indeed, there are at least two instances in which the notions detach from each other.\footnote{For field theories: in gauge-fixing  in the presence of boundaries, seen in section \ref{sec:gf}; and in the computation of Noether charges, in section \ref{sec:hor_charges}.}  Both of these instances require the consideration of bounded regions. 
 
  This touches on one of the  main points I want to bring to the fore of philosophy of physics with this work: the importance of treating bounded regions in our aim to  understand the true nature of gauge systems. Hitherto,  discussions of gauge have steered away from questions of regionality. But it is only once we consider the infinite-dimensional fiber bundles appropriate to the treatment of field theories, and the matter of gluing together the corresponding regional mathematical structures, that we will see just how indispensable gauge degrees of freedom truly are. 

Gauge degrees of freedom are often introduced for practical purposes. For example: the totality of gauge-invariant degrees of freedom cannot be localized, as evidenced by Wilson loops.\footnote{Using Wilson loops as basic variables is also in practice problematic for a variety of reasons \cite{Healey_book}.} The lack of a local description of  all the gauge-invariant degrees of freedom compels us to use a particular (but arbitrary) perspective. In other words, for some types of system, there is no \textit{tractable} `god's eye view'---unveiling all the physics without requiring a perspective.

Still, retaining all these perspectives to describe the same system seems  uneconomical.   Gauge-fixings are usually taken to  rid our descriptions of the superfluosness in a tractable, albeit undemocratic, manner. 
Fixing a gauge amounts to choosing a selective class of perspectives whose members are asked to satisfy certain conditions. Starting from an arbitrary perspective, the conditions should be just strong enough to allow for a unique second perspective, related to the first by a change of frame, to satisfy these conditions and thereby be included in the selective class.\footnote{In standard jargon, each gauge orbit must intersect a gauge-fixing section only once.}  If two gauge-fixed perspectives differ, they \textit{really are} descriptions of physically distinct states; they really do concern different states of affairs. 
Gauge-fixing thus restricts us to perspectives that use the same language to describe their world. Gauge-fixing avoids the comparison of ``apples to oranges'', as the saying goes; using it we compare oranges to oranges and protons to protons.\footnote{ This is in reference to the (approximate) isospin symmetry. This feature of gauge-fixings is well-known in the standard local gauge theory case, but it also applies to  the colloquial use of `perspectives' to discuss vision: fixing one's representation of, for example, a desk lamp in terms of characteristics of the lamp---e.g. `the chord is in direct view'---enables one to unambiguously tell lamps apart \cite{VanFraasen}.} 

Therefore,  one might think, if we can always fully gauge-fix, the plurality of perspectives is surely  superfluous; a surplus structure that we can shave off by quotienting or by agreeing on a common language \cite{Dewar_gauge}. But  a fly lands in this ointment once we countenance  a merely provincial access to the World.

\subsubsection{\textit{Rovelli's relationalism: ``Why gauge'' } --}
The diluted physical meaning usually attributed to pure gauge degrees of freedom is a consequence of the monopoly of physical meaning that is usually awarded to gauge-invariant objects. 

In  \cite{RovelliGauge2013},  a dissonant view  is put forward. It is a fundamentally relational view, and attributes an important role to perspectives. It implies that when we are in possession of only partial subsystems---and therefore lacking the complete set of relations physically characterizing the entire system--- it makes sense to keep gauge-variant information in-hand. 

For Rovelli, gauge-variant\footnote{`Gauge-variant' is prefearble  to the double-negative `gauge non-invariant', as Rovelli and others refer to the  property.} objects are necessary to couple certain types of subsystems.  As he points out, \lq\lq{}gauge non-invariant quantities [...] represent, in a sense, handles through which systems can couple.\rq\rq{} 
To illustrate the idea, he describes two squadrons, each made up of $N$ spaceships, which are just coming into contact with each other. {\,Given  $q_j^1$ as the position of the j-th ship in the first squadron, the Lagrangian  for the first squadron is given by 
\be\label{eq:squadron_L}L_1=\frac12\sum_{i=1}^{N-1}(\dot q^1_{i+1}-\dot q^1_i)^2
\ee
which has  a time-dependent displacement symmetry acting as $\delta_1  q^1_i:= q^1_i+ f_1(t)$  (mutatis mutandis for the second squadron, with `2' in place of `1'). For each squadron we can find gauge-invariant variables by taking the difference in position of two ships $\bar q^1_i=q^1_{i+1}-q^1_i, \,\, i=1,\cdots N-1$ where the barred variables, $\bar q^1_i$ are gauge-invariant. Although there are $N$ spaceships in each squadron, we can describe the subsystems individually in a gauge-invariant manner by using the gauge-fixed $N-1$ variables; of course, we could have chosen many different parametrizations of the gauge-invariant variables. }

{\,Now the second squadron appears and fighting begins. Interaction terms between the two squadrons might no longer be independently gauge-invariant under $\delta_1, \delta_2$. They still could be, if expressed in terms of the barred variables, but for the example Rovelli gives, 
$$ L_{\mbox{\tiny int}}=\frac12(\dot q^1_1-\dot q^2_N)^2
$$ they are not, and invariance requires $f_1(t)=f_2(t)$; only global time-dependent displacements act as a symmetry. There are now $2N-1$ relative distances, and therefore the number of degrees of freedom for the coupled system exceeds the sum of degrees of freedom from the individual systems by one: $\#(1\cup 2)-(\#1+\#2)=1$.} Gauge-invariant degrees of freedom cross the subsystem boundary,  e.g. $q^1_1-q_N^2$, and we  are unable to describe the coupling of the two squadrons merely by using the relational gauge invariant variables of each subsystem; we require their gauge variant  degrees of freedom for coupling. 
 
  Rovelli's succinct explanation for  the existence and importance of gauge-variant variables  follows: 
\begin{quote} ``The gauge invariance of [one squadron] is invariance under an arbitrary time-dependent displacement of all spaceships. The position of the individual ships is redundant in the theory insofar as we measure only relative distances among these. But in the physical world, each ship has a position nevertheless: this becomes meaningful with respect to one additional ship, as this appears. [this additional ship plays the role of a coupling between the two squadrons]. In other words, \textit{the existence of a gauge [symmetry] expresses the fact that a reference is needed to measure the position of a ship. It expresses the fact that we measure position relationally.} It suggests that this is the general case for all gauge systems. The fact that the world is well described by gauge theories expresses the fact that the quantities we deal with in the world are generally quantities that pertain to relations between different parts of the world, that is, which are defined across subsystems. The example shows how a gauge quantity typically describes an individual component of relative observables.'' [my italics]\end{quote}

To cement the ingredients we will need from Rovelli's discussion, let me offer another, more `modern' example. Suppose Alice describes regional spacetime events in her Lorentz frame. She has fixed Lorentz invariance---that ambiguity  no longer acts on \textit{her own frame\rq{}s} description of events. The gauge has been fixed, and a gauge-fixed system is a gauge-invariant system. If Alice\rq{}s region comprises the entire spacetime, we lose nothing by shedding information characterizing Alice's frame itself. Her coordinate description of the Universe encodes all the gauge-invariant physics and we can forget about Lorentz invariance.\footnote{This does not endow this  frame with a privileged status, a point emphasized by  John Bell  \cite{Bell1976}: ``we need not accept Lorentz’s philosophy [of the ether] to accept a Lorentzian pedagogy. Its special merit is to \textit{drive home the lesson that the laws of physics in any one reference frame account for all physical phenomena.}”[my italics]} For a single region, nothing is lost by gauge-fixing.

   But, if the region does \textit{not} comprise the entire Universe,  we must relate Alice\rq{}s description of her region to Bob\rq{}s description of his. To accomplish this, we need to keep some information characterizing the perspectives themselves. Rovelli's example at least recognizes the importance of this information.

Because we have no local description of the physical degrees of freedom of gauge theories, we use particular (but arbitrary) perspectives. But the multitude of perspectives is usually seen as superfluous, since the assumption is that we can always gauge-fix to a single selective class (and so secure a gauge-invariant description). However, without global access to the system, i.e. when we know ourselves to be  provincial in the World, gauge-fixing is premature. Instead, we need to stay flexible about perspectives, which entails keeping gauge-variant elements in the theory. 

\subsubsection{\textit{The language edict } --}
As an analogy, imagine a given region in the World, where all languages are spoken.  As with gauge redundancy, different languages offer different signs to describe the same physical object, and therefore to compare distinct objects it is simpler to use  the same language. Recognizing this,  an edict is announced which obliges the population to speak only Brazilian Portuguese. Since inter-lingual dictionaries there are no longer useful, they are all destroyed, and the inhabitants forget all other languages. 

Later on, this region comes into contact with another region where a similar edict allows only German to be spoken. How do the inhabitants of the two regions now compare objects? In the analogy, a gauge-transformation is effected by inter-language dictionaries, and a gauge-fixing is the settling on one language, once and for all. Thus a regional gauge-fixing  abjures the possibility of a larger World with other languages, rendering dictionaries  entirely redundant. 
But if we don't have the full picture---if there is or could be other regions---we need to keep some flexibility of language, some language-variance if you will. In other words: just as the lack of a tractable `god's-eye-view'  of  the system obliges us to start using perspectives, the lack of a global view  obliges us to keep them. 
 
In sum: Rovelli implicitly attacks a naive approach to all  theories with local gauge symmetry. According to this naive approach, gauge degrees of freedom are at bottom just redundancies in our descriptions. We only need these descriptive redundancies so as to write the theory locally, but then we cull them by fixing the gauge, and in the end a gauge-fixed theory carries just as much ontological content as the non-fixed one. All agree that no objections to this approach arise, \textit{if} we have access to the entire Universe. But they do arise, if we don't have such access.

\subsection{A multitude of perspectives, not a position in the world}\label{sec:contraRovelli}

   
  Although Rovelli has identified one aspect of the importance of gauge-variance which we will later exploit,  Rovelli's construction is  not fully perspectival in the sense that I want to implement here. For note that he requires the `bare position'  of one spaceship in each of the two squadrons---a gauge-variant quantity in each subsystem--- and this only insofar `as the other squadron appears'. These two gauge-variant elements are then combined together, forming one more gauge-invariant quantity. But what if we had chosen a different basis as the gauge-invariant variables of each squadron? Or what if instead of the `bare position' of a ship, we would like to use a different gauge-variant parameter for the first squadron, e.g.: the center of mass of the squadron? Or if the interaction term had been different?  Without  knowledge of which kind of second subsystem will appear over the horizon, it could be a non-trivial task to couple the two new parameters in a gauge-invariant fashion; the two choices must be not only gauge-variant, but gauge-covariant in the right manner. 
  
 Apparently oblivious to this complication,  Rovelli justifies the retention of the gauge-variant variable by resorting to our intuitions about the `real world':  ``But in the physical world each ship has a position
nevertheless: this becomes meaningful with respect to
one additional ship, as this appears.'' But why single out a `bare position'  of a single ship as the gauge-variant handle? How to choose which ship? And can these choices only make sense `as the other squadron appears over the horizon'? After the second squadron has appeared we can in any case, at least in Rovelli's example, describe everything in terms of gauge-invariant observables for the entire system. 

No: the handle must be, as it were, flexible---able to accommodate multiple choices of parametrization---and it must be ready  beforehand, ready to be ``grabbed by'' any \textit{potential} second system.\footnote{This is reminiscent of a discussion of charts and atlases in the characterization of differentiable manifolds \cite{Lang}; and this line of thought can be pursued to reach  conclusions encompassed by ours \cite{DonnellyFreidel, GomesRiello2016},  by analogously keeping, in addition to  the chart description,  the information of the embedding maps of the chart into the manifold (see also \cite{Brown_Kuchar, Isham_embedding, Bergmann_Komar}).  Explicitly: this parametrization, generalizing Rovelli's `position in the real world', would correspond to an embedding map of a chart. In other words, in the differential geometry language, one could argue that if a  coordinate description of the real world $M$ is complete, i.e. there is a global chart, whose inverse we write as $\phi:U\subset \bb R^n\rightarrow M$ such that $\phi(U)= M$, then this description in $U\subset \bb R^n$ exhausts the ontological content of the real world, and we lose nothing by confining our talk of reality to elements of $U$. But if $\phi(U)\neq M$, then we must confer reality on the abstract manifold $M$, in which the domain of the chart is  embedded. In this case, the embedding map itself is significant irrespective of whether there is another intersecting chart. Nonetheless, Rovelli's position is tenable to the extent that specific properties of $\phi$ are only relevant once we are required to use transition functions, i.e.: once there is an intersection with another chart. Then, the ``position of the ship in the real world''  (i.e. the embedding of $U$) becomes indispensable, for it needs to be compared to another perspective (whereby we obtain the transition $(\phi')^{-1}\circ\phi:\phi^{-1}(\phi'(U')\cap\phi(U))\subset U\rightarrow U'$). This distinction---between the reality of the manifold $M$ vs. the reality of transition maps---mirrors that between objective realism and Rovelli's  relationalism. Rovelli's view is part of a larger theoretical framework with which he approaches quantum mechanics, i.e. ``relational quantum mechanics'', in which a realistic interpretation of quantum subsystems only makes sense in relation to other subsystems \cite{RovelliRQM}. }
Thus I will here propose a more general framework, where such a  handle automatically embodies covariance,  without recourse to  the bare position of a ship or to the nature of a second subsystem.

The tenor of the construction is simple. First, we will keep the description of the system through an arbitrary frame; we will keep all perspectives. Then, for any given process, we will extricate the physical component of this process from the effects of a change of perspective. This can be done with respect to each perspective.  Although both the physical and the gauge components of the process are gauge-covariant, one component has a bijective correspondence to the gauge-invariant degrees of freedom of the region, and the gauge component tells us how to couple to other regions.
 We will go through the corresponding mathematics shortly, in the field-theoretic context. 

Accordingly, in this paper we will jettison knowledge about the `bare location' of one ship with respect to another squadron, in favor of maintaining arbitrary perspectives. Accommodating this plurality of views---different descriptions of the subsystems by arbitrary frames---moves our discussion to a big, extremely redundant configuration space, usually described by principal fiber bundles. 
For our purposes here, a principal fiber bundle is essentially a space in which all the different descriptions of the same physical configuration  organize themselves tidily in  subspaces, called `the fibers'. Taking the equivalence class of each fiber would give you a (purely abstract) representation of the physical content of that configuration---the `view from nowhere'. The space of all equivalence classes is usually called `base space', and in many field-theoretic cases it is  an unwieldy ``cubist'' landscape, in which no standard notion of localization  is possible.

 \subsection{Gauge-fixing and boundaries in field theory}\label{sec:gf}
 Apart from its undemocratic nature, gauge-fixing is seen by many to resolve issues of redundancy with gauge. In this subsection, our exploration of gauge field theory in a bounded region will dispel the absolute powers of gauge-fixing, or at least fill them with nuance. 
 
 \subsubsection{\textit{Gauge-fixing in field theory without boundaries}--}
We ended the last section with a brief mention of the sort of spaces that accommodate a multitude of perspectives, generally referred to as principal fiber bundles. As evidenced by Earman's quote \cite{Earman_ode}, the mathematical theory of principal fiber bundles is a staple of gauge theories, be they theories of finite-dimensional systems (like Rovelli's squadron) or more realistic field-theories (like the standard model of particle physics).  The  discussion above was focused on Rovelli's finite-dimensional case. Let us now extend it to the field-theoretic case.\footnote{For the issues that arise in the extension of Rovelli's argument to the Yang-Mills case, see \cite{Teh2015}. Teh discusses the two sorts of couplings  for field-theory which we will cover: between different fields and between different regions. }  

In field-theory, a gauge-fixing is perfectly analogous to the  fixing of one frame of reference as discussed above. For simplicity, we start this exposition in the absence of boundaries, i.e. when our patch of the Universe is the whole Universe. {So here we are assuming a single chart is able to cover the entire manifold.}

 To recap, the `view from nowhere'  is incapable of  \textit{locally} describing the full relational content of the system, and therefore we must start from  an arbitrary perspective. For instance, in Yang-Mills theory, in the principal fiber bundle language, we can only describe fields over spacetime using a \textit{finite-dimensional} `section' (see the left hand side of figure \ref{fig1} for an illustration, which distinguishes the finite-dimensional section from the infinite-dimensional one, representing a gauge-fixing)--- we usually denote this by $A_\mu^a$ (with $\mu$ describing spacetime indices and $a$ the Lie-algebra indices). 

 Since we now have all these arbitrary perspectives, we resort to a gauge-fixing. A gauge-fixing should allow us to directly compare the physics of two states,  originally in arbitrary perspectives, by transforming them into descriptions (perspectives) in the same selective class.  
 A gauge-fixing chooses a selective class of perspectives (or a gauge) by ``intransigently demanding'' that the field configurations satisfy its conditions. 
 
 Such conditions should be, in a sense,  \textit{physical}. What I mean by this is that. they should constrain the frame by the way the system looks in that frame, i.e. they should  constrain the perspective.  For example, in Newtonian mechanics, this could be given by inertial frames,  center of mass, and some anisotropy parameters. In the context of electromagnetism, such a condition could be that the electromagnetic potentials of the configuration in the chosen frame have no temporal component $A^{\text f}_0=0$ and are transverse (i.e. have polarization orthogonal to momenta) $\partial^i A^{\text f}_i=0$ (with script f standing for `fixed'; it is not a Lie-algebra index, which is absent in electromagnetism). In sum, the condition should be a condition  \textit{on the perspective}, not the frame.\footnote{ Indeed, as we will later see, since there is no canonical isomorphism between the gauge group and an orbit of the field under gauge transformations. We cannot implement conditions to fix the frame itself---the field-content must always mediate our choice of frame.  }
 
  Concretely: starting with any potential $A_\mu$, the gauge-fixing procedure  requires us to ``rotate'' our frame (or ``translate'' your gauge potential) until its condition is met. For example: suppose we are in a spacetime of Euclidean signature, then if the gauge-group acts  as $A_\mu\rightarrow ~ A^\lambda_\mu:=A_\mu+\partial_\mu \lambda$ for $\lambda$ an element of the group, one such gauge-fixing condition would require us to solve
 \be \label{Lorentz_gf}
\text f(A):= \partial^\mu( A^\lambda_\mu)=\partial^\mu(A_\mu+\partial_\mu \lambda)=0\Rightarrow \nabla^2\lambda=-\partial^\mu A_\mu
 \ee for $\lambda$, obtaining the field-\textit{dependent} $\lambda^{\text f}(A)$. In the Yang-Mills case, this is known as the Landau gauge (or the Lorenz gauge in Lorentzian signature). Our gauge-fixed field is then the package ${A^{\text f}}_\mu:= A_\mu+\partial_\mu \lambda^{\text f}(A)$. {\,When there are no boundaries, such a gauge-fixing procedure can eliminate all redundancy, allowing  access to the ``gauge-invariant''  degrees of freedom. Equivalently, in the absence of boundaries, trying to fit a different gauge-related configuration, $(A^{\text f})^\lambda$ in the same selective class fails, because  
 \be\label{gf_uni}\partial^\mu ((A^{\text f}_\mu)^\lambda)=\nabla^2\lambda=0 \qquad \text{iff} \qquad \lambda=\text{const}.\ee}
 and a constant gauge transformation doesn't change $A_\mu$.  
  Having got what it wanted, the gauge-fixing immediately forgets---a gauge-fixing maps two field configurations which differ by a frame (or gauge) transformation  to the same representative configuration. 
  
 To sum up: when there are no boundaries, gauge-fixings erase the information pertaining to the arbitrary choice of frames, and thus do not leave gauge-variant handles available for describing coupling. The gauge-fixed theory remembers only a gauge-invariant  package $A^{\text f}_\mu$, and forgets both $A_\mu$ and  $\lambda^{\text f}(A)$. The latter transforms covariantly as $\lambda^{\text f}(A^{\lambda'})=\lambda^{\text f}(A)-\lambda'$. { Therefore given two gauge-fixings, $\text f_1(A)=0,\, \text f_2(A)=0$, the difference 
 \be\label{eq:gluing_simple}F(\text f_1, \text f_2)(A):=\lambda^{\text f_1}(A)-\lambda^{\text f_2}(A)\ee
  is thus a gauge-covariant,\footnote{Only in this simple Abelian case  is it actually gauge-invariant.} group-valued  functional of the perspectives; it relates the two selective perspectives. As we will see in section \ref{sec:hor_symp_geom}, $\lambda^{\text f}(A)$ is an elementary precursor to the object we will use to distinguish changes of frame from physical processes: the \textit{relational connection-form}.   Apart from selecting a representation, the connection-form will be used for relating those representations, as in \eqref{eq:gluing_simple}; the physical representation will be analogous to $A^{\text f}_\mu$ and the gluing will use the analogue of $F(\text f_1, \text f_2)(A)$. 
  
   If the main theme of this paper is the role of the  connection-form  in the treatment of gauge theories in bounded regions, the most prominent sub-theme is that   an approach to gauge systems characterized by perspectives can be pried  away from an approach to gauge systems characterized by abstract (Lie) groups and algebras. The difference between selecting a perspective and selecting a gauge parameter in bounded regions, which we will now witness, manifests this sub-theme. }
  
 \subsubsection{\textit{Gauge-fixing in field theory with boundaries}--}\label{sec:gf_bdary}

  This gauge-fixing-induced amnesia causes no trouble  if we have access to the Universe as a whole; for example,  if Rovelli's world had a single squadron. Nonetheless, the discussion of the last section, conducted in the absence of boundaries, is also germane when they are present.

  Given  two regions $M_1, M_2\subset M$ which share a boundary, $\partial M$, we have the respective restrictions of the $A_\mu$ field onto the regions, $A_\mu^1, A_\mu^2$.   For bounded regions,  finding the perspectives  $A_\mu^{1}$ (resp. $A_\mu^{2}$) which belong to the selective class---satisfying \eqref{Lorentz_gf}---also requires us to stipulate boundary conditions on the \textit{gauge-parameter} in that region, $\lambda^{1}$ (resp. $\lambda^{2}$), itself. These are \textit{not} conditions on perspectives. Moreover, the resulting change of frame---e.g.: the solution $\lambda^{{1\text f}}$---carries a dependence on boundary conditions  throughout the region, and not just at the boundary, as we will shortly see.\footnote{It is also important to note that a Hamiltonian approach would be plagued by precisely the same issues. There, a gauge-fixing is also a function of the canonical variables, $\phi(p,q)=0$, and requires further, non-physical, determination at the boundaries. }

 {   As opposed to what is implied by \eqref{gf_uni}, in the presence of boundaries two arbitrary perspectives which are related by a change of frame can correspond to selective perspectives which are  different everywhere. This is easy to see: Again, we start with  $A^{\text f}$ satisfying the constraints on perspectives, and ask if a transformed perspective, $(A^{\text f})^\lambda$, would be sent back to the same $A^{\text f}$ by the same gauge-fixing constraint which $A^{\text f}$ satisfies, e.g. $\partial^\mu A_\mu=0$. But here  equation \eqref{gf_uni}  leads to $\nabla^ 2\lambda=0$, whose solution is not  $\lambda\equiv$const. for $\lambda_{|\partial M}\neq0$,\footnote{ The solution is a harmonic function which depends on the value of $\lambda$  at the boundary. For example,  according to a mean value theorem for $\nabla^2\lambda=0$ with a prescribed $\lambda$ at $\partial M_1$, for a domain $M_1\subset \bb R^n$, and $\lambda\in C^2(M_1)$, assuming $M_1$ to be a ball centered at $x$ with radius $R$, then $\lambda(x)=\frac{1}{n \text{Vol}(B^n)R^{n-1}}\oint_{\partial M}\lambda(y) dy$ where $\text{Vol}(B^n)$ is the volume of the n-dimensional unit ball. This  illustrates the `percolation' of boundary gauge transformations into the determination of the frame in the interior of the region.}  as opposed to the general solution in the absence of boundaries, \eqref{gf_uni}.  
 
 The possibility of  non-zero solutions for $\lambda$ in this circumstance  evinces the strange character of gauge-transformations at the boundary: different choices of $\lambda$ at the boundary seem to correspond to actually \textit{physically distinct, and yet gauge-related}, configurations. In other words, we face an anxious question: Should we take those perspectives which are gauge-related and yet are not mapped to the same gauge-fixed perspective  to represent physically distinct states of affairs?
  
  The appearance of $\lambda$  at the boundary is therefore far from innocuous; its presence is problematic from two related angles. First,  as described above, it may license physical status to gauge degrees of freedom at the boundary, and second,  fixing $\lambda$ on each side incurs the ``language translation'' issue we saw in the previous section. 
  
   To see the second issue more clearly here, even if we choose the same gauge-fixing (e.g. $\partial^\mu A_\mu=0$) on each side of the boundary, i.e. on $M_1$ and $M_2$,  but  different $\lambda_1, \lambda_2$, it follows that ${{A^{{1\text f}}_\mu}}$  and ${{A^{{2\text f}}_\mu}}$ would not smoothly join at $\partial M$ (they would not even be $C^0$ there, since they differ by  $\lambda_1-\lambda_2$, which is not fixed by the gauge-fixing).\footnote{ Although  we are here focusing on the single chart domain, this discrepancy is perhaps best visualized in  the case of two intersecting charts. Let $M_1\cap M_2=M_{12}\subset M$ be itself an n-dimensional submanifold of $M$. While it is true that, over $M_{12}$, one demands $A_\mu^1=(A_\mu^2)^\lambda$ for some gauge transformation $\lambda:M_{12}\rightarrow \bb R$, the gauge-fixings of regions $M_1$ and $M_2$ could be different, and there are no gauge-transformations to relate the two so selected perspectives. Even more worrisome,  due to the non-local nature of gauge-fixing, even if one has chosen the same constraint on the partial perspectives, e.g. $\partial^\mu A_\mu^1=\partial^\mu A_\mu^2=0$, the different domains in which these equations are solved (and in particular the different boundary conditions on $\lambda^1, \lambda^2$), imply that in general  ${{A^{{1\text f}}_\mu}}\neq {{A^{{2\text f}}_\mu}}$ on $M_{12}$. So even if they speak the same language they can't communicate!\label{ftnt:overlap}} Alas, there are no more gauge transformations available; the gauge-fixing on each side has left no gauge-covariant handles, and we can no longer `rotate' ${{A^{{1\text f}}_\mu}}$  and ${{A^{{2\text f}}_\mu}}$ to match. This can be seen as a field-theoretic analogue of both the metaphor of the two regional languages incapable of translation, and Rovelli's  obstruction to the description of the coupling between the gauge-fixed squadrons.\footnote{
 In Rovelli's case, when  $\dot q_1^1=\dot q^2_N$,  the coupling is no longer problematic. In the field-theory case, there might likewise be particular coincidences,  $\lambda_1=\lambda_2$, for which the gauge-fixed configurations match at the boundary, but neither of these cases is generic and so do not resolve the problem. The circumstance of this coincidence will be clarified in due course (section \ref{sec:gluing}). \label{ftnt:coincidence}} 

Here we see that the boundary choices for $\lambda$, being non-perspectival and gauge-variant, create puzzles.  One could attempt to resolve these puzzles   by disallowing gauge transformations at the boundary.\footnote{This is related to the algebraic approach to the issue of boundary degrees of freedom, for which a vast literature exists, especially in the context of entanglement entropy, see e.g.: \cite{Casini, Harlow}. The values of the physical fields at the boundary are also fixed in many of these approaches.\label{ftnt:Casini}} Even disregarding the many technical questions that still remain in this case,  I do not think that the approach is conceptually satisfactory. After all, nothing is explained about the different character of gauge degrees of freedom at the boundary. 

 But we could also aim to rectify these puzzles by demanding that variations of the boundary conditions on $\lambda$, i.e. $\delta\lambda_{|\partial M}\neq 0$,  still yield the same physical degrees of freedom. As we will see  in section \ref{sec:connection_form}, allowing for such variations requires the introduction of  the connection-form in field-space. Incidentally, the connection-form employs  a unified bulk and boundary  gauge-fixing of sorts, i.e. if translated to the present context,  its boundary conditions would be of the form $\text{f}(A_{|\partial M})=0$, gauge transforming covariantly. 

 \subsubsection{\textit{Boundary degrees of freedom?}--}
Before ending this section, enabled by the  the previous discussion on boundary gauge-fixings,  let us take a look at the motivation for the introduction of actual physical degrees of freedom at the boundary.

 Supporting this line of argument, one could point out that in field-theory, the space of global physical degrees of freedom does not nicely decompose into regions. Indeed,  one can imagine  Wilson loops for non-Abelian theories, which cross the boundaries, and thus support  gauge-invariant functions only on the joint region, being gauge-variant when restricted to each region (see figure \ref{fig:region}).\footnote{For Abelian theories, this argument doesn't exactly go through. One may always decompose a loop intersecting the boundary between the two regions into two loops, which coincide at the boundary but run in opposite directions. In Abelian theory, the observable obtained by the total loop can be recovered from the two regional observables. In non-Abelian gauge theory, one needs to take traces of holonomies, and the observables therefore don't compose in the same way. }
\begin{figure}[t]
		\begin{center}
			\includegraphics[scale=0.17]{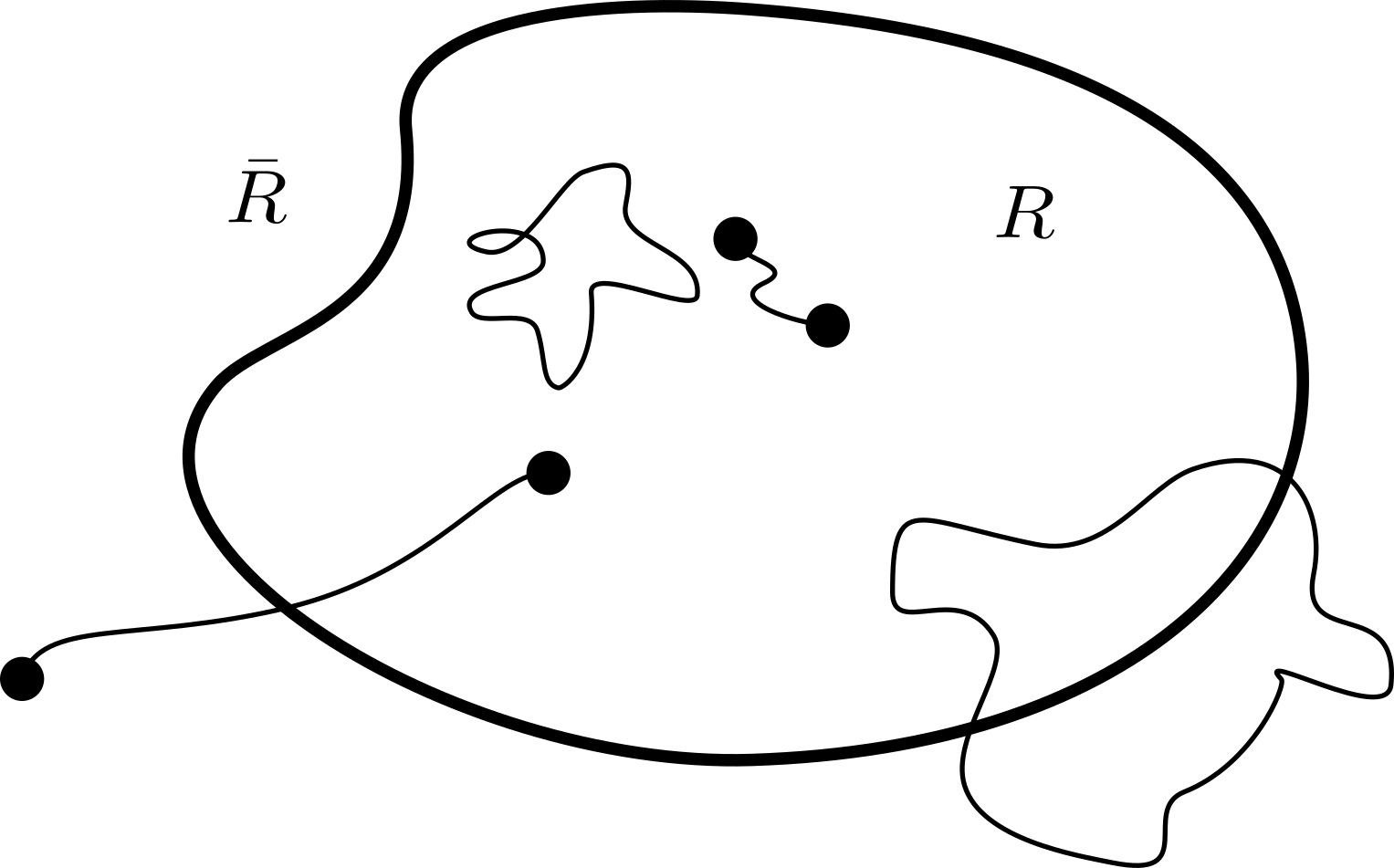}	
			\caption{ A region and a selection of loops both contained and not contained in it. There exist gauge-invariant functions whose support is not contained in either the inside or the outside of the region. Therefore the space of physical, gauge-invariant functions does not factorize, as the quotient of field space itself does not factorize: $\F/\G\neq \F_R/\G_R\otimes \F_{\bar R}/\G_{\bar R} $ where $\F$ is the field space associated to the entire region, and its restrictions are denoted by a subscript (and we are supposing each quotient is inheriting a vector space structure, for which we can form the tensor product).}
			\label{fig:region}
		\end{center}
	\end{figure}
 
 Strangely, upon gluing two regions back together without the prior regional gauge-fixing, the choice of $\lambda$ at the boundary becomes completely immaterial.  If $\G_{1}$ is the group of gauge-transformations in $M_1$, and $\G_1^o$ are those that reduce to the identity at $\partial M$, i.e. 
 \be\label{phys_edge} \G_1^o:=\{g\in \G_1\, |\, g(x)=\text{Id}\,\,\text{ for} \,\,x\in \partial M\}\ee  we seem to have added physical degrees of freedom in correspondence to $\G_{1}/\G^o_1$---for each physical degree of freedom,\footnote{One could be tempted here to refer to such degrees of freedom as ``bulk'' degrees of freedom. But I will refrain from doing this, remaining agnostic about localization. } a different element of $\G_{1}/\G^o_1$ would correspond to a new physical degree of freedom as well---which must disappear once the regions are glued back together!\footnote{ It is important to note here that such cutting and gluing is here a purely abstract operation. This does sound a lot like the job description of ghosts in non-Abelian QFT. We have clarified this analogy in \cite{GomesRiello2016} and I will comment on it further along the paper. }  


If one decides to embrace the physical status of boundary gauge degrees of freedom, one should consistenly promote the boundary gauge parameter required for the gauge-fixing to the same status as the original fields.  Furhter, one should  imbue these new degrees of freedom with  covariance properties under changes of gauge, so that the coupling of two regions becomes gauge-invariant (thus fulfilling the relational role advocated by Rovelli). 
This is essentially what Freidel and Donnelly have accomplished in \cite{DonnellyFreidel} (see also  \cite{Donnelly_entanglement, ReggeTeitelboim1974, Brown_Kuchar, Isham_embedding, Bergmann_Komar} for related approaches and precursors). Their language and purpose was different than the one we have espoused here; as we will discuss in section \ref{sec:hist_intro}, their primary aim was   to obtain gauge-invariant Noether charges (to which we turn in the next section).  For that, they introduced new degrees of freedom into the theory which appear only at the boundaries of subsystems.  In other words, these degrees of freedom are kept in reserve, to be pressed into battle only  when our region is confronted with another, when they can  be used for coupling. 

	But, if such degrees of freedom exist, a host of questions would need to be pursued. How seriously should we regard the boundary as possessing its own physical degrees of freedom, as real as any other? Can we detect them? How do we see them only at abstract boundaries?  As we will see in section \ref{sec:hor_charges}, using the connection-form as opposed to edge-modes, all of these puzzles evaporate. 
	
	The question that we turn to next, in the way of explicating the nature of the boundary gauge degrees of freedom and the origin of Donnelly and Freidel's edge-modes, is: if these were true degrees of freedom, would they carry their own charges?

 \subsection{Boundaries and charges: a brief intro.} \label{sec:hist_intro}

{\, In the first subsection, \ref{sec:Rovelli}, I have argued that the nature of gauge degrees of freedom is  `perspectival'. Following Rovelli's example, I argued that  in the presence of subsystems this view  underlines the significance of  gauge-variant objects, to the detriment of gauge-fixed quantities. Then, in the second subsection, \ref{sec:contraRovelli}, I argued contra Rovelli:  keeping the flexibility required to  couple to different subsystems, and also required to support different gauge-fixings (different choices of gauge-invariant variables), demands a more formal, pluripotent structure than the one Rovelli offers (``a position in the World''). These were the perspectives. The challenge is to extract physical information while keeping a multitude of perspectives in play; this will be the job of the connection-form, as we will see from section \ref{sec:varpi_abstract}  onward. 
 
{\,Indeed, in the third subsection, \ref{sec:gf}, we saw that the usual procedure of gauge-fixing in bounded regions still suffers from  the problem Rovelli identified: coupling between different regions is still hampered. There is again a mismatch in the  counting of the physical degrees of freedom belonging to the parts and belonging to  the whole. In the field-theoretic context, this mismatch can be associated to an artificial specification of gauge-parameters at the boundary.\footnote{By ``artificial'' I mean it is unrelated to conditions $\text{f}(A)=0$ through which perspectives can be selected.}}

 Its consequences are not merely conceptual, but can be felt ``on the ground'' in the physics of charges.   For even if we grant that the spirit of gauge degrees of freedom is relational, only to be fully seen in the coupling of subsystems, their bite comes from the teeth of the Noether theorems. And the unphysical parameters at the boundaries are implicated in the complicated procedure of obtaining physical charges from \textit{local} gauge symmetries through Noether theorems.

\subsubsection{\textit{ The teeth of gauge: Noether theorems} -- }

The first Noether theorem  implies that global  symmetries result in conservation laws that we observe empirically (see \cite{BradingBrown, BradingCastellani} and references therein for a historical account and philosophical discussion). The conservation of electric charge, for example, is a consequence of global $U(1)$ symmetry, conservation of energy  in special relativity is a consequence of time translation symmetry, etc. 

 Whereas global symmetries are usually understood to possess empirical significance, for instance in the form of conserved charges,  
  the empirical significance of local gauge symmetries is much looser. For these symmetries are in the domain of Noether's \textit{second} theorem, whose effect is better construed as constraining the form of the equations of motion, rather than as straightforward conservation of charge.\footnote{For instance, in the case of the internal groups of Yang-Mills, they will yield Bianchi identities for the curvature tensor; while  for diffeomorphisms, they will yield generalized Bianchi identities for metric concomitants (see \cite{HenneauxTeitelboim} chapters 1-3, and \cite{BradingCastellani} for a philosophical discussion). In the Hamiltonian language, they give us the local constraints and their closed algebra. It is easy to see in a particular example that indeed they can yield conserved charges for special gauge-parameters (related to Barnich and Brandt's ``reducibility parameters'' \cite{BarnichBrandt2003}. For GR, the Bianchi identities imply the conservation law for energy and momentum, $\nabla_aT^{ab}=0$. Integrating it against a Killing vector, $\xi^a$ such that $\nabla_{(a}\xi_{b)}=0$, we obtain, for $\epsilon, \epsilon_a$ the oriented volume densities of $M$ and $\partial M$ respectively: $$\int_M \epsilon\, \xi_b \nabla_aT^{ab}=0=\oint_{\partial M} \epsilon_a \xi_b T^{ab}$$}

   Explicitily,  the  puzzlement about conservation laws comes into focus  in the distinction between Noether's first and second theorem. As two of the foremost experts in the topic characterize the issue: \cite{BarnichBrandt2003}:
\begin{quote}
``[The puzzle] is encountered when one tries to define the charge related to a
gauge symmetry “in the usual manner”, by applying Noether’s first theorem on the
relation of symmetries and conserved currents. The problem of such an approach is that
a Noether current associated to a gauge symmetry necessarily vanishes on-shell (i.e., for
every solution of the Euler-Lagrange equations of motion), up to the divergence of an
arbitrary superpotential.''
\end{quote} 
{ Gauge charges are related to constraints on initial data, at Cauchy surfaces. Therefore, at least in the absence of boundaries of these surfaces, such charges should vanish when the constraints are obeyed \cite{HenneauxTeitelboim}. 

Following the standard methods by which one can associate  charges to symmetry-generators in a covariant setting, these so-called Noether charges indeed vanish in the absence of spatial boundaries.\footnote{The standard  method I refer to here is the covariant symplectic framework, which will be more fully explained in section \ref{sec:hor_charges}. Here I aim to merely illustrate the problem.} However, this is not the case for bounded subregions of  the Cauchy surface. 

Let us take  electromagnetism as an example.  We define the electric field as a d-2 form (for d the dimension of $M$, the  spacetime)  $E=\ast F$, where $F$ is the electromagnetic curvature tensor and $\ast:\Lambda^n(M)\rightarrow \Lambda^{d-n}(M)$ is the Hodge star operator, taking n-forms (i.e. elements of the alternating tensor product of linear functions on $TM$, denoted by $\Lambda^n(M)$) to the complementary $d-n$ forms, $\Lambda^{d-n}(M)$. Then, let  $R\subset \Sigma\subset M$, bounded by $\partial R$, be a region of the Cauchy surface $\Sigma$. For a gauge-generator $\xi(x)$, the covariant  procedure (which we explain in section \ref{sec:hor_charges}) yields a Noether charge for this field content and gauge parameter:
 \be\label{superpotential_charge}
 Q[\xi] \approx\oint_{\partial R} \left(\text{Tr}(E(x)\xi(x))+\alpha[\partial_\mu \xi]\right)d^{d-2}x,
 \ee
 where  $\approx$ indicates equality up to terms that vanish when the equations of motion (including the constraints) are satisfied, also called ``on-shell equality''; and $\alpha$ is an \textit{arbitrary} (field-independent) $\Lambda^{d-2}(M)$-valued  linear functional of $\partial_\mu \xi$; in the language we will introduce in section \ref{sec:hor_charges}, $\alpha\in \Lambda^{d-2}(M)\otimes\Lambda^1(\F)$ ( $\Lambda^1(\F)$ means it is a linear functional on field-space, cf. section \ref{sec:hor_geom}).   The point is that such charges are doubly troubled: not only do they seem to be non-zero for those gauge-generators which do not vanish on the boundary, but the arbitrariness of $\alpha$  even makes it hard to define a specific charge for any given transformation!
 
 }
 
Therefore, although talk of conservation at codimension one spatial surfaces (codimension two in spacetime) might sound a lot like a Gauss law, there are important differences:  these quantities exist at the corners of the space-like Cauchy surfaces bounding a given region, more conditions are required to discuss their evolution; and they may be infinite in number, and do not solely depend on the original physical fields and geometrical shape of the d-2 surface $\pp R$ in \eqref{superpotential_charge}. The integrated  charges from \eqref{superpotential_charge} will be smeared by the generators of arbitrary gauge-transformations at the boundaries, and may depend on arbitrary choices of superpotentials, which strains physical  interpretations of such charges.

Naively calculating charges for continuous symmetries by Noether's second theorem therefore depends on arbitrary  choices  at the boundary. According to the arguments from the previous section, this arbitrariness at least partially corresponds to the one we saw in the gauge-fixing procedure.\footnote{One might think that one could use the rigid asymptotic symmetries to obtain the actual charges at infinity. But this is also more complicated than it seems. As Barnich and Brandt assessed the situation in 2003 \cite{BarnichBrandt2003},  
``The problem of defining and constructing asymptotically conserved currents
and charges and of establishing their correspondence with asymptotic symmetries \textit {in a
manifestly covariant} way has received of lot of attention for quite some time."[my italics]  And that interest has only grown in the past years (see \cite{strominger2018lectures} and references therein).} 

\subsubsection{\textit{Gauge-variant `edge-modes'} --} 

Taking  Regge and Teitelboim's seminal work on conserved charges at asymptotic infinity  \cite{ReggeTeitelboim1974} as a guide,\footnote{To my knowledge, Regge and Teitelboim were the first to introduce new physical ``embedding variables'' at boundaries to reinstate a broken symmetry (in their case, Poincar\' e). But there are many more, very closely related precursors to Donnelly and Freidel. Most notably, \cite{Brown_Kuchar, Isham_embedding, Bergmann_Komar}, have all introduced degrees of freedom to parametrize gauge choices. }   Donnelly  and Freidel  introduced  a new special type of gauge-variant degree of freedom \cite{DonnellyFreidel}.  It  encodes intrinsic boundary degrees of freedom, for boundaries both asymptotic and finite, and  either real or imagined. 

The manner in which Donnelly and Freidel  design the gauge-covariance properties of the boundary degrees of freedom in \eqref{surface_gauge} is such that  their effect on the charges will end up canceling the first term of \eqref{superpotential_charge}.\footnote{ In fact, the terms are added at the level of the symplectic potential, and not  directly at  the level of the action nor of the charge.  Moreover, their introduction is predicated on the \text{field-dependence} of the gauge-parameters at the boundary. This is less of a stretch of the usual concept than might at first seem, for, as we saw in section \ref{sec:gf}, boundary gauge transformations have a ``physical''  character. In other words, changing the gauge at the boundary implies a physical change of the field. \label{ftnt:DF_deltag}.}
But if this were the end of the story, Donnelly and Freidel might have had a problem:  this procedure would identically cancel all charges. That is where the following equation \eqref{surface_symmetries} comes in.

Calling the gauge-group-valued new degree of freedom $\kappa(x)\in G$ for $x\in \partial M$, with  $A$ for the Yang-Mills gauge potential, $\xi$ the generator of an arbitrary gauge transformation, $\D=\d+[A,\cdot]$ the gauge-covariant derivative (with the Lie-algebra bracket $[\cdot, \cdot]$ and $\d$ the exterior derivative) and $\delta_\xi$ indicating the infinitesimal action of the gauge-transformation on the variables,   they define: 
\begin{align}
\text{\bf Gauge symmetry:}&\qquad \delta_\xi\kappa:=\kappa \xi, \qquad \delta_\xi A:=\D\xi \label{surface_gauge}\\
\text{\bf Surface symmetry:}&\qquad \Delta_\beta\kappa:=\beta\kappa,\qquad \Delta_\beta A:=0\label{surface_symmetries}
\end{align}
where $\beta$ are the new generators of symmetry of $\kappa$, and I have suppressed: (i) indices (notation  will be more thoroughly introduced in equation \eqref{eq_infin_gtransf}) and 
(ii) the specific form of the left and right action of the Lie algebra on the surface degrees of freedom. 
 The \textit{surface symmetries} generated by $\Delta_\beta$ are supposed to represent redefinitions solely of the boundary degrees of freedom. These redefinitions are cordoned to only include the original gauge group, i.e. $\beta\in \G_{\partial R}$. Other than repackaging the original symmetries, the motivation for this choice is nebulous to me.  

  
   

 In other words, Donnelly and Freidel are able to precisely exploit the arbitrariness in the definition of $\alpha$  in  \eqref{superpotential_charge} so as to cancel the charges associated to $\xi$. The covariant properties of these   new boundary degrees of freedom are designed to render the charges $Q$ completely gauge-invariant. For the action of the gauge parameters given in \eqref{surface_gauge}, no charges are left. However,    these edge modes also bring their own infinite contributions to charges; namely all the charges associated to the action of the gauge-parameters through $\Delta_\xi$, given in the second line, \eqref{surface_symmetries}.  In the end,  the tally is unchanged: there remain a continuous number of left-over `physical'  degrees of freedom associated even to imaginary finite boundaries, and these yield the gauge-invariant charges.

 In my view, starting from the field content and the boundaries of a given region, we would like  charges to be entirely functionally dependent on field content and geometrical features of the boundaries---and not at all dependent on any further arbitrary local gauge choices. As with \cite{BarnichBrandt2003}, we should then expect charges to be strictly associated to \textit{physically relevant symmetries}, such as Killing transformations in the case of general relativity and non-Abelian Yang-Mills.\footnote{Or, more precisely: strictly related to reducibility parameters \cite{BarnichBrandt2003}.} But  Donnelly and Freidel's  formalism does not solve this issue.\footnote{But they also do not hang much on the issue. In the words of Donnelly, ``I am comfortable with the point that our charges depend on new degrees of freedom. And indeed, those are not there when the regions are combined [...] And I also agree that one could try to build such degrees of freedom out of fields already in the theory, which is an appealing idea.'' Private communication. }

To sum up, the  reason to deem Donnelly and Freidel's new charges unphysical is  simply this: their charges  depend on a new field which has no physical interpretation if boundaries are not present. In my personal view,  this approach may cause some confusion if boundaries are not material entities, but just figments of the theorist's imagination.  Looking ahead, the point is that both imbuing physical significance to the left-over gauge choice $\lambda$ at the boundary, and making up new degrees of freedom are unnecessary for solving any of the issues considered above: there is another way!  


 Up to this point of the paper,  we have essentially reviewed some of the vexing issues arising from conjoining  gauge and boundaries, and considered one attempt at resolution \cite{DonnellyFreidel}. Such issues are technical and conceptual, and must be embraced and dealt with by all who worry about the nature of gauge. 
 
  Now,   we  arrive at a different attempt at resolution, one which seems to patch up the holes left by the first part of the paper. 
I will give a conceptual exposition of \cite{GomesRiello2016, GomesRiello2018, GomesHopfRiello}. Instead of new degrees of freedom held in reserve at the borders, ready to be pressed into duty, we use \textit{relational properties of the original field-content}. These are present everywhere within the regions, and  are also able to perform the special translation functions at the boundaries, for their covariance is a consequence of their geometric nature.




\section{Horizontal  geometry and the connection-form}\label{sec:hor_symp_geom}
We now turn the spotlight to the main actor of this paper:  the \textit{relational connection-form}. Given our  purposes here, the following sections will consist of an extremely abridged account of the work contained in \cite{GomesRiello2016, GomesRiello2018, GomesHopfRiello}, without the technical detail that can be found there.  The first subsection, \ref{sec:summary_varpi}, summarizes the role and construction of the connection-form. 
{  The second subsection, \ref{sec:hor_geom},  puts up the  technical scaffolding,  by establishing some required notation and concepts and giving an overview of the horizontal geometry of field-space.  The third subsection, \ref{sec:varpi_abstract}, then gives an abstract mathematical definition of the connection-form.  }

\subsection{A summary of the construction}\label{sec:summary_varpi}

 The  connection-form resolves the problems listed above by a much simpler route---it defines physical and pure gauge transformations intrinsically to each perspective (e.g.: to each $A_\mu$) without having to  introduce spurious degrees of freedom  \cite{gomes_riem, GomesRiello2016, GomesRiello2018, GomesHopfRiello}.  It embodies the function of `Rovelli's gauge-variant couplings' by describing the system from a given arbitrary perspective, while keeping the required gauge-variant handles characterizing that perspective. 
 
 Let us stress two main differences between a connection-form and a gauge-fixing: as I have repeatedly emphasized,   gauge-fixings  are amnesiac---they forget information about the frame and therefore do not leave gauge-variant handles around to describe coupling between subsystems.

A second difference is that a gauge-fixing is in the business of \textit{defining a selective class of perspectives}, while a connection-form is in the business of \textit{defining changes in perspective}.  That is, given any field configuration $\varphi$, a gauge-fixing eats it up and spits out a  $\varphi^{\text f}$, which is a  state gauge-equivalent to $\varphi$  but with a different perspective, viz. one belonging to the selective class. 

Here is what the relational connection-form does instead:  given the arbitrary state of affairs $\varphi$ and an infinitesimal transformation of that state, 
$\bb X_\varphi$ (or $\delta \varphi$ in the usual notation), the connection-form decomposes that transformation into a pure gauge translation---the change of frame---and a \textit{physical} change. This change is physical with respect to $\varpi$ and with respect to that particular field-content of $\varphi$; it depends on the perspectives in a covariant manner, as evidenced by equation \eqref{eq_fundamental}, below.
 The pure gauge part is what provides the handles for coupling at the boundaries as we will see in section \ref{sec:gluing};   there are no new degrees of freedom that make curious special appearances there.

 These advantageous properties can be directly inherited from the geometry of field space, as we will see more directly in section \ref{sec:examples}. In essence, this is because the connection-form rectifies the other issues identified at the end of section \ref{sec:gf}. Namely, requiring covariance also under different boundary conditions,  $\delta \lambda_{|\partial R}\neq 0$, requires the use of a connection-form covariant under variations, $\delta$ (and not only under derivatives, $\partial_\mu$), i.e. \textit{a field-space connection-form}. In turn, using such a connection-form and unrestricted gauge transformations at the boundary, we recover a sort of boundary expression of gauge-fixing which is covariant and of apiece with the bulk gauge-fixing condition, i.e. something of the sort:  $\text{f}(A_{|\partial M})=0$.
 
  Due to its geometrical representation, the connection-form  defines a modified, horizontal symplectic geometry in field-space. This sort of horizontal geometry is germane to the distinction between physical and pure gauge (vanishing) Noether charges,  more so  than  standard symplectic geometry, even if the difference is only explicit at corners and boundaries. 
  Indeed, as we will see in section \ref{sec:hor_charges}, the connection-form is much more conservative on the topic of Noether charges than Donnelly and Freidel's boundary degrees of freedom.\footnote{Although it has not yet been done explicitily, it is expected that the results for entanglement entropy in gauge theories obtained by the procedure of Agarwal et al \cite{Agarwal} are entirely reproduced by the connection-form.\label{ftnt:Agarwal}}

\subsection{The geometry of field-space}\label{sec:hor_geom}

The more explicit construction relies on geometrical structures  of field-space, to which we will now turn. 


 In the forthcoming, $\F$ is  to be thought of as the space of all possible field configurations ---all the possible states of affairs of the field, with no regard to satisfying the equations of motion.  This vastly redundant space represents all of the possible `perspectives' on the same state of affairs, as discussed in section \ref{sec:intro}. Local gauge transformations act locally on  the fields in question, relating the perspectives.  The action of the group forms ``fibers'' in  this space, partioning the entire field-space into different equivalence classes. 

Such a fibration looks a lot like the standard principal fiber bundles we usually encounter in gauge theories. There, base space is usually just spacetime. However, here, in this infinite-dimensional field-theoretical context, due to a lack of local parametrization of all the local gauge-invariant degrees of freedom, base space can only be characterized as the quotient $\F/\G$, ``the moduli space of  physical field configurations''. 

In the case of gauge potentials, the moduli would be obtained by setting $A' \sim A\Leftrightarrow A'=R_g A$ for some right action of the gauge group ($R_g$) on the potential, and we would denote the equivalence classes by $[A]=[A']\in \F/\G$. For diffeomorphisms acting on spatial metrics, elements of $\F/\G$ would be geometries---i.e each \textit{indivisible point} of $\F/\G$ would consist of the complete geometric structure of space, without redundancy. As far as the `view from nowhere'  can be implemented,  it applies only to $\F/\G$.\footnote{ {Each point of $\F/\G$ characterizes a full gauge-invariant configuration. However, gauge-invariant observables of GR are known to be non-local, e.g. integrals over spacetime such as $\int R^{\mu\nu}R_{\mu\nu}\sqrt{g}$, and therefore each such point in $\F/\G$ cannot be put into correspondence with a region or point of $M$.} There can be no surprise appearances of Einstein's `hole argument' in $\F/\G$, since each point of it has  identified all diffeomorphically-related metrics. However, there are several problems about endowing a differentiable local orbifold structure on  the quotient space of Lorentzian-signature spacetime metrics \cite{isenberg1982slice}. But it is straightforward for the space of Euclidean signature metrics of any dimension.}

Apart from the status of the base space, there is another important disanalogy between standard finite-dimensional principal fiber bundles (PFBs) and $\F$  seen as a fibered space over $\F/\G$. As we will see in section \ref{sec:hor_charges},  it is crucial that the latter is  \textit{not} a bona-fide PFB. That is because not all the orbits in $\F$ are isomorphic---some are effectively of lower dimension than others. This structure emerges from differences in frame which do not amount to differences in perspective. These differences carve out the landscape features of $\F/\G$ which will give rise to global charges. The fact that such a structure manifests itself directly in  the physical moduli space is what we would expect if such charges have  physical content.

 Nonetheless, as in the case of finite-dimensional PFBs, it is useful to define a notion of `sameness of frame'  when moving from one orbit to another on $\F$---this is the geometrical role of the connection-form.  Note that, unlike a gauge-fixing, this notion of sameness is only defined infinitesimally, and conforms to different perspectives of the same configuration (i.e. is covariant). 

\subsubsection{\textit{Mathematical preliminaries} -- }
More specifically, the stage on which we set the pieces is field-space, denoted by  $\F=\{\varphi^I\}$.
In this notation, $\varphi^I$ stands for a whole field configuration $\{\varphi^I(x)\}_{x\in M}$, where $I$ is a super-index labeling both the field's type and its various components, and $M$ denotes the underlying manifold (space or spacetime).

	In the following, a `double-struck' typeface---as in $ \dd$,   $\fLie$, $\fvf$, etc.---will be consistently used for field-space entities. For instance, we introduce on $\F$ the deRham differential $\dd$ \cite{ Crnkovic:1986ex, Crnkovic:1986be, Crnkovic:1987tz}; it should be thought of as the analogue, on $\F$, of the standard spacetime differential (or exterior derivative) $\d$. We will also need a notation for field-space vectors,  Lie derivatives and the interior product between forms and vectors in field-space, denoted respectively by $\bb X$,  $\fLie$ and $\fI$. For instance, contraction of a vector  with  a basis element of the differential forms in $\F$, denoted by $\dd \varphi^ I$, is defined by
	\be
	\fI_\fvf \dd \varphi^I = \fvf^I \qquad \text{where} \qquad \bb X=\int\, \bb X^I\frac{\dd}{\dd \varphi ^I}
	\ee
	where I omitted dependence on $x$ in the integrand for simpler notation. 
\begin{figure}[t]
		\begin{center}
			\includegraphics[scale=0.17]{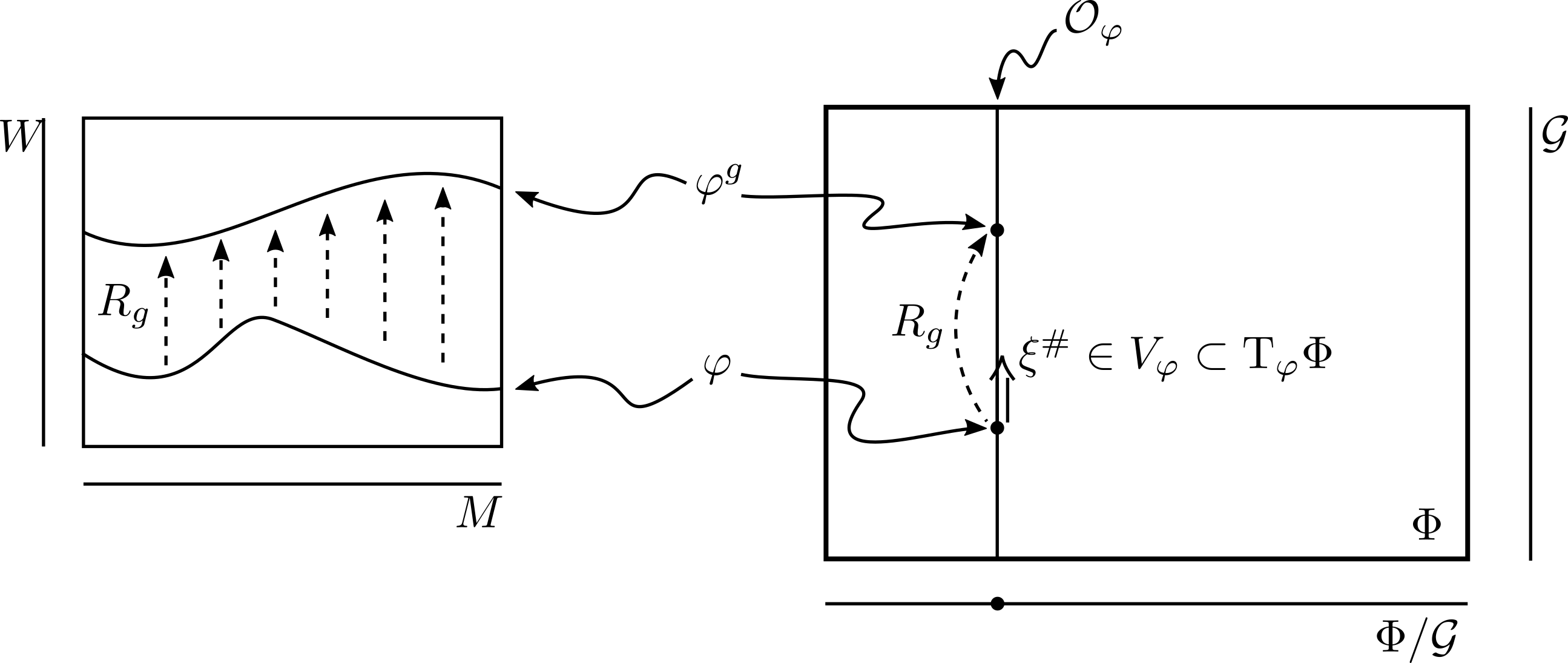}	
			\caption{The field-space $\F$ seen as a principal fiber bundle.  $\F/\G$ stands for the quotient space of `gauge-invariant configurations'.  I have highlighted a configuration $\varphi$---it is one perspective on $[\varphi]\in \F/\G$, which is intractably non-local,---its (gauge-transformed) image under the action of $R_g:\varphi\mapsto\varphi^g$, and its orbit $\mathcal O_\varphi \cong \G$.   On the left hand side of the picture, we see a representation of $\varphi$ and $\varphi^g$ as sections of a vector bundle over the spacetime region $M$. }
			\label{fig1}
		\end{center}
	\end{figure}
		
Also for simplicity, we will focus on an internal gauge group, but at this level there would be no difference between this action and, say, an action of diffeomorphisms on the space of metrics  (it still acts pointwise on field-space \cite{gomes_riem, GomesRiello2016}). Given a \textit{charge group}, say $G=SU(N)$, gauge transformations also inherit a group structure, forming the \textit{gauge group}:
	\be
	 \G := C^\infty(M, G),
	\ee
	$\G$ with elements $g(\cdot) \in \G$ and point-wise composition (we are now moving beyond the Abelian case) over $M$, and $(\cdot)$ denoting the slot for points of $M$.  Similarly, the Lie algebra of the gauge group is given by
	$
	\xi(\cdot) \in \fG = C^\infty(M,\fg)
	$ 	where $\fg={\rm Lie}(G)$.  There is a right action of the group:
	 \be
	\begin{array}{rccl}
		R :&  \G\times \F &\to& \F  \\
		&\Big(g(\cdot), \varphi\Big) &\mapsto& R_{g(\cdot)}\varphi=:\varphi^g.
	\end{array}
	\label{eq_right_action}
	\ee
	Given this right action, we  define a map from the Lie algebra of the gauge group, $\fG$, into the vector fields on field-space (denoted by $\fX(\F)$),
	\be
	\begin{array}{crcclc}
		&{}^\# :&  \fG &\to& \fX(\F)  &\\
		&&\xi &\mapsto& \xi^\# &
	\end{array}
	\label{eq_sharpoperatordef}
	\ee
	This map associates the  flow $\xi^\#\in\fX(\F)$ on field-space to an infinitesimal gauge transformation $\xi(\cdot)\in\fG$. 
	 Such $\xi^\#$ are called {\it fundamental vector fields}, formally defined iteratively from its action on a scalar $f:\F\rightarrow \bb R$:
	 \be
	{\xi}^\#f   := \fLie_{\xi^\#}f := \frac{\d}{\d t}_{|t=0} f(R_{\exp(t{\xi})} A, R_{\exp(t{\xi})}\Psi) .
	\ee
where we have already used  the field-space of Yang-Mills theory, $\FYM$,  given by the two different sectors--- gauge connections, $A$, and matter fields, $\Psi$---i.e.:
	\be
	\FYM = \{\varphi =  (A,\Psi) \} .
	\ee
	In this context, part of the physical field content is given by the standard $\fg$-valued 1-forms over the spacetime manifold
	\be
	A = A^a_\mu(x) \tau_a\d x^\mu \in \Lambda^1(M, \fg),
	\ee
	where $\fg= \text{Lie}(G)$ and $\{ \tau_a \}_a$ is an orthogonal basis of $\fg$.
	We take this basis to be normalized with respect to the trace (in the fundamental representation) as $\tr(\tau_a \tau_b) = -\frac12 \delta_{ab}$.

	We will consider here only scalar  matter fields, $\psi$: smooth functions on $M$ valued in $W$, where  $W$ is a vector space carrying the fundamental representation of $G$, e.g.: for $W=\bb C^N$ and $G=SU(N)$:\footnote{This restriction is only for simplicity in exposition. In \cite{GomesHopfRiello}  we consider also spinorial fields.}
	\be
	\psi = \psi^m(x)|m\rangle \in \mathcal{C}^{\infty}(M,W);
	\ee
	where $\{|m\rangle\}_{m=1,\cdots, N}$ is a basis for $W$. 
	
	The transformation properties of $A$ and $\psi$ are given by:
	\be
	A^g = g^{-1} A g + g^{-1}\d g 
	\quad\text{and}\quad
	\Psi^g = g^{-1}\Psi.
	\label{group_action}
	\ee
	so we obtain
	\be
	\xi^\# = \int \delta_\xi A \frac{\dd}{\dd A} + \int \delta_\xi \Psi \frac{\dd}{\dd \Psi}
	\label{xi_hash}\ee
	where we  introduced the standard notation for infinitesimal gauge-transformations (along $\xi$),
	\be
	\delta_\xi A = \D \xi := \d \xi + [A,\xi]
	\qquad\text{and}\qquad
	\delta_\xi\Psi = - \xi \Psi,
	\label{eq_infin_gtransf}
	\ee
	with $[\cdot,\cdot] $ the Lie bracket on $\fg = {\rm Lie}(G)$, extended pointwise on $M$ to $\fG$.

	Gauge orbits in $\F$ are ``canonical''  (there is no extra choice to be made), and so is their tangent space,  called `vertical'. The vector fields $\fvf$ that are tangent to the orbit at a given  $\varphi\in\F$ define through their span a vertical subspace $V_\varphi$ of the tangent space $T_\varphi\F$:
	\be\label{eq:ver_space}
	{\quad\phantom{\Big|}
		\mathrm T_\varphi \F \supset {V}_\varphi = \mathrm{Span}\{{\lvf}^\#, {\lvf}\in\fG\}.
		\quad}
	\ee
	Vertical fields represent infinitesimal  gauge transformations---infinitesimal changes of perspective---and they span the gauge orbit through a given point. 	To briefly illustrate our notation and construction so far, consider figure \ref{fig1}. As shown in the figure, we call an orbit through a point $\varphi$, $\mathcal{O}_\varphi$. 
	
	Now,  $[\varphi]\in \F/\G$ is defined abstractly by taking the equivalence class, and there is in general no natural parametrization for the gauge-invariant degrees of freedom.  $\varphi$ represents one perspective, or one parametrization of $[\varphi]$. But, since there is no canonical isomorphism between $\G$ and $\mathcal O_\varphi$, a given $\varphi\in \mathcal O_\varphi$ does not itself single out a frame. However, for each $T_\varphi\F$, we seem to have a canonical isomorphism between $\fG$ and $V_\varphi$, i.e. we can canonically identify a change of a frame (i.e. $\xi\in \fG$)  to a change of perspective (i.e. ${\lvf}^\#=v\in V_\varphi$). 
	 In the nomenclature we have been using, this is what \eqref{eq_sharpoperatordef} does.

But in fact, this identification may falter: although for standard principal fiber bundles, $V_\varphi$ and $\fG$ are isomorphic vector spaces, this is not always the case for the infinite-dimensional, field-theory cases. There are changes of frame which do \textit{not} bring about changes of perspective. We can say such changes are in the \textit{blind-spot} of the given perspective. 

As we shall see in section \ref{sec:hor_charges}, the implicit assumption that the isomorphism above always holds spawns much confusion regarding the distinction between global and local symmetries. 
As we envisaged in section \ref{sec:Rovelli}, the failure of this isomorphism illustrates the second way  in which the standard characterization of gauge systems through their symmetry group detaches from that given by our intuitions on perspective.\footnote{The first concerned the stipulation of the gauge transformations at boundaries, for gauge-fixing.} 

\subsection{Connection-forms: the formal construction}\label{sec:varpi_abstract}

Now, figure \ref{fig1} looks like a principal fiber bundle. As with any principal fiber bundle, we should aim to define a connection-form therein, which tells us how to parallel-propagate frames from orbit to orbit.

Indeed, such a definition is equivalent to defining a horizontal complement to the vertical subspaces in the tangent bundle $T\F$.
	Emulating the finite-dimensional case \cite{kobayashivol1},  $\varpi$ (pronounced VAR-PIE) is defined as a functional 1-form over field-space, valued in the Lie algebra of the gauge group $\fG$,
	\be\label{map:varpi}
	\varpi \in \Lambda^1(\F, \fG).
	\ee
	A one-form naturally contracts with vector fields, and thus  its kernel defines some distribution (see figure \ref{fig2}). If we moreover demand that the connection give a bijection between the vertical space and the Lie algebra, i.e.: 
	\be\fI_{\xi^\#}\varpi  = \xi  \label{varpi_compl_0}\ee
	then the kernel defines a  horizontal distribution:\footnote{At least in finite dimensions, \eqref{varpi_compl_0} sums up \eqref{eq:hor_space} and \eqref{eq:ver_space}. In infinite dimensions some further conditions need to be fulfilled. See e.g.: \cite{gomes_riem} and references therein.}
	\be\label{eq:hor_space}
	{\quad\phantom{\Big|}
		H := {\rm ker}(\varpi) = \{ \fvf \in \mathrm T\F \,|\, \fI_\fvf \varpi = 0\}.
		\quad}
	\ee
	\begin{figure}
		\begin{center}
			\includegraphics[scale=.17]{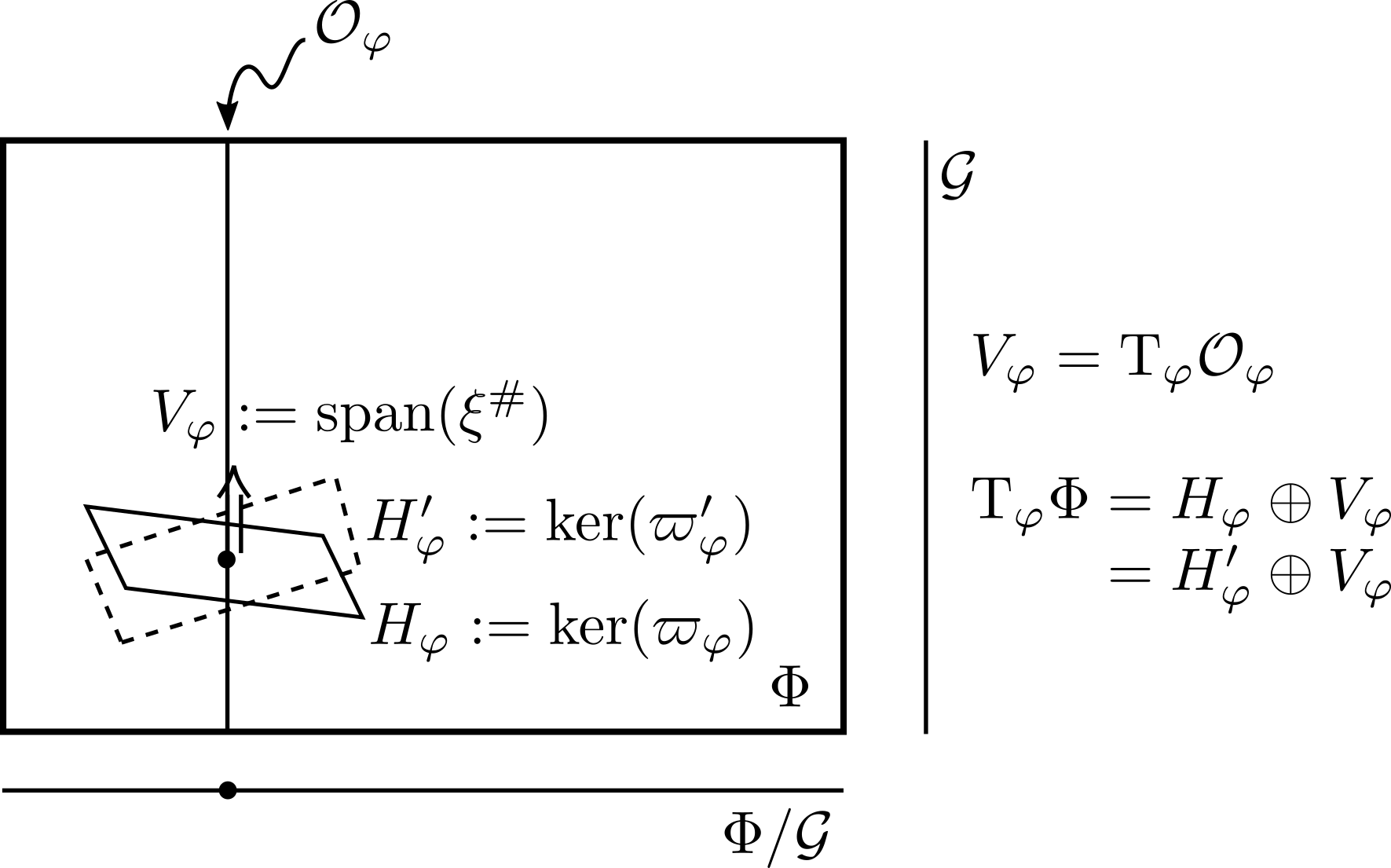}
			\caption{The split of ${\rm T}_\varphi \F$ into a vertical subspace $V_\varphi$ (that spanned by $\{\xi_\varphi^\#, \xi\in\fG\}$) and two different choices for horizontal complement $H_\varphi, H_\varphi'$ defined as the kernel at $\varphi$ of a functional connection $\varpi$ and $\varpi'$ respectively.  Although each choice of functional connection $\varpi$ defines a distribution of subspaces of the tangent bundle, this distribution is not necessarily integrable, i.e. not necessarily tangent to `horizontal submanifolds' . }
			\label{fig2}
		\end{center}
	\end{figure}
Using this notation, we can thereafter identify the verticall projector by $\hat V = \varpi^\#$, i.e.: since $\fI_{\bb X}\varpi\in \fG$, for $\bb X\in T_\varphi\F$,  we have $\hat V(\bb X)=(\fI_{\bb X}\varpi)^\#\in V_\varphi$ and therefore the horizontal projection is defined by  
\be \label{eq:hor_proj}
\hat H = ({\rm Id} - \varpi^\#)\,:\, \bb X\mapsto \hat{H}(\bb X)=\bb X-(\fI_{\bb X}\varpi)^\#\ee
 The horizontal exterior derivative is obtained by composition with the horizontal projection \cite{kobayashivol1}: 
\be\label{hor_ext_der} \dd_H:=\dd(\hat{H}(\cdot)).
\ee
	
	Now, how  \textit{does} one define such a $\varpi$, precisely? We will shortly see an explicit example. Before that, to be compatible with the gauge structure, a bona-fide connection-form in a principal fiber bundle needs to satisfy the following equivariance condition:
	\be \fLie_{\xi^\#}\varpi = [\varpi,\xi] +\dd \xi \qquad \label{varpi_dependent_0}\ee
	Putting the two equations, \eqref{varpi_compl_0} and \eqref{varpi_dependent_0}, required for the definition of a connection-form together for convenience: 
	\begin{subequations}
		\begin{empheq}{alignat=1}
		\fI_{\xi^\#}\varpi & = \xi  \label{varpi_compl}\\
		\qquad \fLie_{\xi^\#}\varpi &= [\varpi,\xi] +\dd \xi \qquad \label{varpi_dependent}
		\end{empheq}\label{eq_fundamental}
	\end{subequations}

	The standard way of automatically implementing  conditions \eqref{eq_fundamental} is to have the horizontal spaces be orthogonal  to the vertical spaces, with respect to  some physically relevant metric on field-space, i.e. with respect to  a \textit{gauge-invariant supermetric}. We will see this in section \ref{sec:examples}. In short, each sector of field-space (i.e. a field $A$ or $\psi$ and region in space or spacetime in $M$) carries an essentially  unique choice of ultralocal supermetric; and so practical choices are more constrained than they might at first appear.\footnote{For Yang-Mills there is a single such supermetric, and for gravity a one-parameter family. Ultralocality---i.e. a metric that is a first integral of undifferentiated field-space vectors--- is required for the well-posedness of properties of the connection, as we will glimpse in section \ref{sec:examples} (see section 5 in \cite{GomesHopfRiello}).  If any disambiguation is necessary, we can also appeal to the kinematical field-space metric which implicitly appears in the Lagrangian of the theory in the 3+1 case. }

	 Therefore, having chosen the sector, and then given the corresponding supermetric, one has a canonical choice of connection. Importantly, as we will see in section \ref{sec:examples}, we can build connection-forms which  only employ the original fields. In this way, it measures what is a physical (horizontal) and what is a gauge (vertical) change  with respect to the instantaneous configurations of those fields, i.e .with respect to what we have called `a perspective'. For these reasons, it is in general called a `\textit{relational connection-form}'. It is relational with respect to a perspective, given by the field content and its  representation in a frame. According  to  \eqref{eq_fundamental},  the vertical (pure gauge) component of a process automatically transforms covariantly, fulfilling the role of `handles'  required by Rovelli \cite{RovelliGauge2013}.

\section{Deploying connection-forms}\label{sec:connection_form}
We are now ready to provide a couple of explicit examples and put  the connection-form to use. 	
We start with the examples in subsection \ref{sec:examples}. In section \ref{sec:gluing}, we analyze how the connection-form translates between physical degrees of freedom of different regions. i.e. how it aids ``gluing''.  Finally, in section \ref{sec:hor_charges}, we show how the horizontal symplectic geometry defined with the aid of the connection-form passes the ``acid test'' of providing physically meaningful charges, without gauge-fixing and without the introduction of extra degrees of freedom.
	 
\subsection{Examples of connection-forms}\label{sec:examples}

{ For Yang-Mills with matter,  the connection for  the  $A^a_\mu$-sector is more specifically termed \textit{the Singer-DeWitt} connection, and for the $\psi$-sector,  \textit{the Higgs} connection. This terminology is  defended in detail in  \cite{GomesHopfRiello}.  Importantly, even for regions with boundary, gauge-transformations are completely unconstrained at the boundary; no conditions apart from the field sector and an appropriate supermetric need to be specified. 

Although we will not see here the explicit equations of the connection for gravity, the general connection's role in selecting preferential frames is best illustrated in examples with non-internal symmetries. Therefore, before we go on to  specific examples of connections, I will first present the  more palpable example of diffeomorphisms so as to demonstrate this role  (see also \cite{gomes_riem} for more technical details). }

\subsubsection{\textit{Selecting frames: the gravity example}--}

 We consider gravity in a 3+1 framework, so that now $M$ is a 3-dimensional manifold representing space, on some definition of  simultaneity (or instantaneous slices). We describe our instantaneous metric by $\{g_{ab}(x), x\in M\}$. The covariant tensor $\mathbf{g}(x)=g_{ab}(x)$ (in abstract index notation) can be subject to active spatial diffeomorphisms (which are the ones we consider), and therefore any such description identifies points in $M$ in a rather arbitrary way, i.e. in an arbitrary ``frame''.\footnote{We cannot identify `coordinate systems' as such, in either the field space, $\F$, or in the gauge group, Diff$(M)$. Each frame here is better thought of as `an identification of spatial points', which can be shuffled by active diffeomorphisms. }

The Lie-algebra associated to the group of diffeomorphisms of $M$ is the set of smooth spatial vector-fields $\fX(M)$.\footnote{ Although we will not concern ourselves with the order of differentiability here,  it does make a difference to the type of manifold structure we endow to both $\F$ and its group action. See \cite{gomes_riem} and references therein for more details. In the $C^\infty$ case, the so-called \textit{inverse limit Hilbert} structure is  the most appropriate \cite{Ebin}.}
A $\varpi$ for $g_{ab}$ and a metric velocity, $\dot g_{ab}$ in this frame, according to \eqref{map:varpi},  we get a spatial vector field $\fI_{\dot{\mathbf{g}}}\varpi^a=:\varpi^a(\dot{\mathbf{g}})\in \fX(M)$. Suppose  $\varpi^a(\dot{\mathbf{g}})=X^a\neq 0$. This vector field, $X^a$, tells us that we changed our frame, or rather, we changed our `identification of points'  during this transformation of $g_{ab}$. It pinpoints which part of the field-transformation that we described as $\dot g_{ab}$ came from a `changing' identification of spatial points along time. In order to say `the identification changed', we need a preferred way of identifying spatial points during evolution.  In the principal fiber bundle language, when integrated over time, $\varpi$  would effect a horizontal lift; it parallel transports a given choice of initial point $x$ along time, providing, in the words of Barbour, an \textit{equilocality relation} \cite{Barbour_Bertotti}.

According to \eqref{varpi_dependent}, this equilocality does not really care about what the original frame was; instead, it only evinces the field's `preferred way' of identifying spatial points amidst an arbitrary change of fields.  

The actual physical change is extricated from the total one by rotating the frame back to  equilocality (and correspondingly adapting an  infinitesimal change in perspective):
\be\label{hor_g}\hat H(\dot{\mathbf{g}})=\dot{\mathbf{g}}
-(\fI_{\dot{\mathbf{g}}}\varpi)^\#=\dot{\mathbf{g}}
-\mathcal{L}_{\mathbf{X}}\mathbf{g}\ee
where $\mathcal{L}_{\mathbf{X}}$ stands for the spatial Lie derivative along $X^a=\fI_{\dot{\mathbf{g}}}\varpi^a$. The properties of  \eqref{varpi_dependent} guarantee that upon a time-dependent diffeomorphism, $\hat H(\dot{\mathbf{g}})$ will have the standard covariance properties under time-independent diffeomorphisms \cite{gomes_riem, GomesRiello2016}.\footnote{Indeed, in the ADM formalism \cite{ADM}, the `shift' components of the metric satisfy the same transformation laws, for field-independent diffeomorphisms.}


It could also be that the transformation involved no physical change at all, and was purely one of  changing perspectives. When that is the case, equation \eqref{varpi_compl}  demands  that $\varpi$ recognizes it to be so, yielding $\dot{\mathbf{g}}=\mathbf{X}^\#:=\mathcal{L}_\mathbf{X}\mathbf{g}$, and therefore, according to \eqref{hor_g}, $\hat H(\dot{\mathbf{g}})=0$.

In the case of diffeomorphisms, differently to previous definitions of equilocality relations \cite{Barbour_Bertotti}, the ones presented here are explicitly gauge-covariant, and are grounded on the field-content of the theory; they can use different fields to inform the choice of frames along time. 

The internal gauge example works almost identically to this diffeomorhism in gravity example, the difference being that  the gauge-potential value is matched along time at predetermined spatial points. 

For matter fields, the situation is similar; here  the main difference, as we will see, is one between \eqref{YM_varpi_boundaries}, which is non-local for both gauge-fields and gravitational fields, and \eqref{varpiSU2}, which is local. In other words,  matter fields define equilocality relations  in a  local manner. Maybe we would like the identification of points along time to be grounded on the location of dust particles around ourselves, or maybe we would like points to be identified by that cloud of neutrinos zipping by---no problem, there will be a connection-form associated to each, as long as these fields are not zero anywhere and have no `blind-spots'.

 `Blind-spots' are infinitesimal gauge-transformations, $\xi$, which are not registered by a given choice of  $\varpi$, i.e.they are  $\xi\neq 0$ such that $\varpi(\xi^\#)=0$. They signal the failing of the isomorphism between changes of frame and changes of perspective: thence their name. That is, a blind-spot is a change of frame which is not registered by the perspective. 
 
 There would be no such $\xi$ if \eqref{varpi_compl} was completely valid, i.e. if $\F$ was a bona-fide principal fiber bundle and $\varpi$ a bona-fide connection-form therein. But they are not. And this failure is immensely important in the determination of Noether charges, as we  will see in section \ref{sec:hor_charges}.

	 \subsubsection{\textit{The general Singer-DeWitt connection }-- }\label{sec:sdw}
	  Let us take a simple, 3+1 Yang-Mills example to understand how $\varpi$ is obtained  through orthogonality: given a Lagrangian for the fields, $L$, and a choice of field-space sector, say $A_\mu^a$, upon a 3+1 split the Lagrangian itself will yield a kinematical supermetric for $A_i^a$ (where now $i,j$ run over spatial indices). In the Yang-Mills case, the kinematic metric is suggested, not surprisingly, by the kinetic term in the Lagrangian: 
$$\bb G(\dot A, \dot A):=\int_\Sigma d^ 3x\,\, \text{tr}\left(g^{ij}\dot A_i^a \dot A_j^b\right)$$
This kinetic metric enables us to define  those instantaneous field-transformations which are  strictly horizontal, since the choice of vertical vectors is canonical  (i.e. of the form $\delta_\xi A=\D\xi$, also called pure-gauge). The definition of these horizontal vectors at each base point in field-space is tantamount to a definition of $\varpi$. It  splits any ``change'' into a purely gauge part, $\hat V(\dot A)$, and a purely physical part, $\hat H(\dot A)$, with respect to the perspective of $A_\mu$ (and what it characterizes as kinetic).

{ To ascertain the precise form of the Singer-DeWitt (SdW) connection, the procedure is simple: first, we determine the set of horizontal vectors, 
$\bb X=\bb h\in H_\varphi:=(V_\varphi)^\perp\subset T_\varphi\F$:
\be G(\xi^\#, \bb h):=\int d^ 3x\,\, \left(g^{ij}\,\bb h_i^a\, \D_j\xi_a\right)=0\qquad \forall \xi\in \fG
\ee 	 
 using the definition of $\D$ in \eqref{eq_infin_gtransf}, and without imposing any a priori restrictions on the content of $\xi$ of $\bb h$ at the boundary of the region $\partial \Sigma$.  Second, we define the horizontal projection, $\bb X^h=\bb X-(\fI_{\bb X}\varpi)^\#$ such that $\bb X^h\in H$.} 
	 In this way, the Singer-deWitt connection is defined as a solution to:
	\be
	{\quad\phantom{\Big|}
		\D^k\D_k \varpi = \D^i \dd A_i
		\quad\text{and}\quad
		n^i\D_i \varpi_{|\pp \Sigma} = n^i\dd A_i{}_{|\pp \Sigma}.
		\quad}
	\label{YM_varpi_boundaries}
	\ee
where $n^i$ is the normal to the boundary. 
	
	In other words, the gauge-covariant Poisson equation for $\varpi$ comes automatically equipped with nonzero \textit{gauge-covariant} Neumann boundary conditions. The appearance of the field-content on the right hand side of the boundary problem is  essential for covariance and tells us $\varpi$ does not represent an arbitrary new degree of freedom encoded in the boundary, as was the case with the arbitraty stipulation of $\lambda$ at the boundary for gauge-fixings, explored in section \ref{sec:gf}.  As predicted in section \ref{sec:gf_bdary}, we implemented covariance wrt varying boundary conditions on the gauge parameter---there denoted by $\delta\lambda_{|\partial M}\neq 0$,--- and simultaneously framed the  (analogue to the)  gauge-fixing in both boundary and bulk in a unified manner, i.e. as (the infinitesimal analogue to) $\text{f}(A)=0$ and $\text{f}(A_{|\partial M})=0$. 
	
	The boundary conditions appearing in the  definition \eqref{YM_varpi_boundaries}  descend directly from the properties of the supermetric in field-space, and therefore are unlike standard boundary conditions, which are restrictions on field content.  Lastly, note that although  $\varpi$  is non-local---as signalled by  inversion of $\D^2$--- these equations are defined intrinsically in each region, from the field content itself, and are for these reasons termed \textit{regional}. 
	
	This connection-form is non-local and, for a non-Abelian charge group, also non-flat---it is non-integrable and therefore does not correspond to any gauge-fixing. At this point of the discussion, it is tempting to say a few words about the Gribov problem. The `problem'  is  more of an obstruction, which tells us that, for non-Abelian gauge theories, $\F$ does not have a global section (or global gauge-fixing). If $\F$ were a principal fiber bundle, the lack of a global section would translate to the non-triviality of the bundle (meaning it does not admit a global product structure). 
	
	A lack of a global gauge-fixing is certain to spell trouble for non-perturbative quantizations of non-Abelian theories, but  let us put that aside for now. The Gribov obstruction tells us that any connection-form which is everywhere defined in $\F$ does not form an integrable distribution, because, if it did, it would be associated to a global section. Since it is not associated to a global section, parallel transport according to $\varpi$ will be path-dependent. That is,  for a non-Abelian gauge theory, $\varpi$ will possess intrinsic curvature, $\bb F$, a very intersting concept on its own. For more on this and its relation to the field-space Aharanov-Bohm effect, see \cite{GomesHopfRiello}. 

	 \subsubsection{\textit{The Abelian Singer-DeWitt connection }-- } 	
	n the Abelian case, i.e. for electromagnetism, the connection is flat. It is given by 
\be
	{\quad\phantom{\Big|}
		 \varpi = \nabla^{-2} (\pp^ i \dd A_i)
		\quad\text{with}\quad
		n^i\partial_i \varpi_{|\pp \Sigma} = n^i\dd A_i{}_{|\pp \Sigma}.
		\quad}
	\label{QED_varpi_boundaries}
	\ee
For $n^i\dd A_i{}_{|\pp \Sigma}=0$, and $\Sigma\simeq \bb R^3$, the unique solution is:
\be\varpi=\int_\Sigma \frac{\d^3 y}{4\pi} \frac{\pp^i\dd A_i(y)}{|x-y|}\ee

	For field-space curves $A_i(x,t)$,  we can define a horizontal lift in analogy to a Wilson line, even  in the general non-Abelian case, through a path-ordered exponential, i.e. the field-dependent field-space \lq{}parallel-transport\rq{}
\be
h_\gamma[A] = \PPexp \int_{\gamma} \varpi\,,
\label{h}
\ee
where $\gamma$ is the history of the field, a path linking the configuration $A$ to the initial configuration $A^\star=A(0)$. 

In the Abelian case, in temporal gauge, we identify $\dd A_i=\dot A_i$ and we can drop the dependence of $h_\gamma$ on the path: $h_\gamma\rightarrow h$. If we set out an initial configuration (such as $A(0)=0$ for electromagnetism), we can construct a unique holonomy associated to each field configuration.  The path integration essentially cancels out with the time derivative (see \cite{GomesRiello2018} for details), and we obtain a ``dressing'':\footnote{The obstruction for such dressings in non-Abelian theories is that they cannot be path (e.g. history)-independent, and are thus not intrinsically defined by instantaneous perspective (the state of the field) \cite{GomesHopfRiello}.} 
	\be\label{Dirac_dressing}
	h[A]:=\int_\Sigma \frac{\d^3 y}{4\pi} \frac{\pp^i A_i(y)}{|x-y|}
	\ee
	
	 In this case, $h[A]$ becomes precisely the gauge-fixing translation $h[A]\equiv\lambda^{\text f}(A)$ discussed in section \ref{sec:gf}, associated to temporal Coulomb gauge.
	Using it, we can  `dress'  variables with a garment that makes them invisible to gauge-transformations. This procedure precisely recovers the results of Dirac \cite{Dirac:1955uv}; using the nomenclature of the introduction, he redefined the electron variables as  $\psi_h:=\lambda^{\text f}{(A)}^{-1}\psi $.   These redefined, dressed variables are gauge-invariant.\footnote{In Dirac's construction it is also important that the electric field created by $h(A)$ is the Coulomb field of the electron. In classical terms, this means that the Poisson bracket of $E$ with $\hat \psi$ is $ \{ E(x) , \hat \psi(y) \} = \frac{-1}{4\pi (x-y)^2}\hat \psi(y)$. Notice that this is a Poisson bracket between gauge-invariant quantities. It is most easily computed in temporal gauge; in Coulomb gauge, in which $h_\text{Dirac}\equiv\rm id$, but one needs to first introduce non-local Dirac brackets. The supermetric appearing in the kinetic term of the Lagrangian governs both the definition of canonical momenta and also the structure of the Gauss constraint. It is not difficult to convince oneself  that the relationship between the SdW connection and the supermetric is what guarantees  that the Poisson bracket $\{E, \hat \psi\}$ gives the expected, Dirac result.}
	
	Moreover, the garment is versatile: it is equally applicable globally and regionally. It is solely defined by the field-content, including its boundary values; one does not freely fix the boundary value of $\lambda$, for here it does not represent a new degree of freedom.

		 \subsubsection{\textit{The Higgs connection} --} 
	The Higgs connection as a concept is perhaps even richer than the Singer-deWitt connection, and, since I  will not be able to convey its many facets here, I must redirect the interested reader to chapter 7 of \cite{GomesHopfRiello}. 
	
	For now, let's take a simple example of Yang-Mills theory where the gauge group acting on the scalar matter fields is $G=\SU(2)$, and 
in this case, $\tau_a = -\frac{i}{2}\sigma_a$. 
We also have a unique type of bilinear term for the fields (see equation 7.3. in \cite{GomesHopfRiello}), and, by constructing the connection-form again by orthogonality, we  obtain  the connection-form:
\be
\varpi_{\rm \SU(2)} 
= \frac{i }{  \Psi^\dagger\Psi}\Big(   (\dd \Psi^\dagger)\sigma^a\Psi - \Psi^\dagger \sigma^a (\dd \Psi)  \Big)\tau_a.
\label{varpiSU2}
\ee 
where $\psi^\dagger$ is the transpose of $\psi$. 
Unlike what is the case for SdW, we have found a local expression for the connection-form. 
The general shape of \eqref{varpiSU2} can be generalized to include more general charge groups, spinors, and also vielbeins transforming under Lorentz symmetries. 

Using an actual matter field allows us to  define the frame point by point. Importantly, this description is only possible for those states of $\Psi$ whose density is everywhere non-zero, i.e. $\Psi^\dagger(x)\Psi(x)\neq 0,\,\, \forall x\in M$.  This caveat has profound implications, and can be related to certain aspects of the Higgs mechanism for symmetry breaking. 

A Higgs connection-form requires a spontaneously broken phase, using the condensate as a gauge-frame, hence the namesake \cite{GomesHopfRiello}.  Coordinatizing the field $\Psi$  as 
\be
\Psi = \Psi(h,\rho) = \rho h v_o
\qquad
\text{where} 
\;
h \in  C^\infty(\Sigma,G)
\;\text{and}\;
\rho \in  C^\infty(\Sigma,\mathbb R),
\label{eqcoord}
\ee
and $v_o\in V$ is some non-vanishing reference vector in $W\cong \bb C^N$, the fundamental representation of $G=\SU(N)$.
Note that  $v_o$ is a reference vector for internal space. It is not a field-space coordinate (i.e. it is independent of the state), hence 
\be
{\quad\phantom{\Big|}
\varpi = - \dd h   h^{-1}.
\quad}
\label{varpim_flat}
\ee
 This connection-form formally recovers  Donnelly and Freidel's edge-modes,  with a simple translation $h_{|\pp \Sigma}\equiv \kappa$ (the edge-modes $\kappa $ appear explicitly in \eqref{surface_gauge},\eqref{surface_symmetries}). The main difference is that here $h$ has a relation to a physical field, $\Psi$, and not just on the boundary. Interestingly, $h$'s physical status brings a caveat: the matter field $\Psi$ cannot vanish at any point within a region for the regional connection-form to exist.

		 \subsubsection{\textit{Barbour-Bertotti's best-matching} --} 
To fully explain the philosophical background to Barbour and Bertotti's best-matching procedure would take us too far afield, as would delving into its consequences for the relationalist vs substantivalist debate (see chapter 4  \cite{Pooley_review}).   The gist of it is depicted in figure \ref{fig:bm} (the idea originated in \cite{Barbour_Bertotti}; see \cite{Flavio_tutorial}, section 6.2, and \cite{Barbour_Mach} for  reviews). This will be a `bare-bones' exposition, based on the concepts we have here developed.

In the language being employed here, a `best-matched' frame can be summarized as a horizontally projected one, when configuration space is the one of N- particles in Newtonian mechanics, $\F=\bb R^{3N}$ and  the gauge group is the 3-dimensional Euclidean one,  $\text{Eucl}(3):=SO(3)\ltimes \bb R^3$, acting as:\footnote{Barbour also considers the so-called `similarity group', composed of  rigid rotations, translations and dilations (an overall change of scale).}
$$q_i^a\mapsto R^a_b\,q_i^b+t^a\qquad \forall i
$$ where $\mathbf{q}_i\in \bb R^3$ is the position vector of the $i$-th particle,  and $(\mathbf{R}, \mathbf{t})\in SO(3)\times \bb R^3$. Finally, to determine the connection-form, we need to stipulate  the field-space inner product, which is derived from: 
$$ d(\mathbf{q}, \mathbf{q}')=\left(\sum_i \|\mathbf{q}_i- \mathbf{q}'_i\|^2 \right)^{1/2}
$$
where  we use the standard Euclidean norm. In this case, the group action is field-independent, and the connection-form would yield the necessary action of the group to attain the ``best-matched'' frame (in the infinitesimal case, i.e. as applied to $\dot {\mathbf{q}}$). 

\begin{figure}[t]
		\begin{center}
			\includegraphics[scale=0.5]{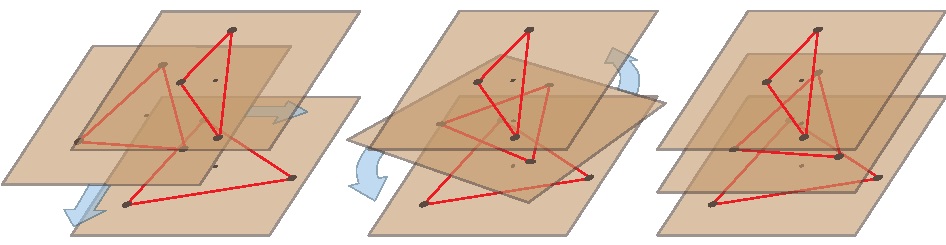}	
			\caption{A stacking of three-body configurations: the arbitrarily chosen stacking above is best matched (blue arrows) by translations so as to bring the barycenters to coincidence, after which rotational best matching eliminates residual arbitrary relative rotation (taken with permission from arxiv  \cite{Flavio_tutorial})
 }
			\label{fig:bm}
		\end{center}
	\end{figure}

\subsection{Gluing}\label{sec:gluing}

 \subsubsection{\textit{Establishing communication: gluing different regions} --}
Now consider a region ${M}={M}_{1}\cup{M}_{2}$, with ${M}_{1, 2}$ embedded manifolds sharing a portion of their boundary, $S=\pp {M}_{1}\cap\pp{M}_{2}  \neq \emptyset$ (with embedding maps $\imath_1:M_1\rightarrow M$,  $\imath_2:M_2\rightarrow M$  $\imath_{S}:S\rightarrow M$). 
	
	To each of these regions, we assign an own field-space, Lie-algebra of gauge transformations, supermetric, and the respective $\varpi$ connection. I will denote a restriction to one of the regions by the same subscripts; that gives us a map $\cdot_J: \F\rightarrow \F_J$ and so on, with $J=1,2$ (i.e. the restriction is of `base-point', given by $\phi(x)\mapsto \phi(x_J)$ for $x\in M$ and $x_J\in M_J$). 
	
	 Since the Higgs matter connection is local, the decomposition into vertical and horizontal follows suit, and there is not much complication at the boundary. Therefore, we will  follow the SdW connection of section \ref{sec:sdw}, but here exploring the consequences solely  for gluing. I.e. we take a connection-form based on \eqref{YM_varpi_boundaries}, reserving a few comments on the Higgs-type connection form to the end of the section.
	
	We can then define the vertical and horizontal projectors $\hat V = \varpi^\#$ and $\hat H = ({\rm id} - \varpi^\#)$, and similarly for $\hat V_{I}$ and $\hat H_{J}$. Each of these operators acts on field-space vectors intrinsic to each region.
	More concretely, given a vector $\bb X$ supported on ${M}$, it can be decomposed into $\bb X = \bb X_{1} + \bb X_{2}$, where $\bb X_{I}$ live respectively on ${M}_{J}$ understood as intrinsic manifolds with boundary.
	
	 While it is true that the  physical (horizontal) component of the restrictions $ \bb X_{1}, \bb X_{2}$ may not match at their boundary, the difference at the boundary is  always purely vertical. But by itself verticality does not suffice to characterize a degree of freedom intrinsic to the boundary; the difference must also be a vertical element \textit{intrinsic to the boundary}, i.e. a field-space vector such that $\bb X^\mu n_\mu=0$ (reinstating the spacetime indices). Indeed, as shown  in equation 5.17 in \cite{GomesHopfRiello} we have: 
	\be\label{bdary_step}\hat{H}_1(\bb X_1)_{|S}-\hat{H}_2(\bb X_2)_{|S}=-(\hat{V}_1(\bb X_1)_{|S}-\hat{V}_2(\bb X_2)_{|S})=\xi_S^{\#_S}(\bb X_1,\bb X_2)
	\ee
where $\bb X_{|S}$ means we restrict $\bb X(x)$ to $x=y\in S$, and where ${\#_S}$ is defined for the field-space and gauge-transformations over $S$, $\F_S$ and  $\xi_S\in  \text{Lie}(\G_S)$, respectively.  This is essentially an  elaborate, infinitesimal, non-Abelian version of the more qualitative \eqref{eq:gluing_simple}. The gauge transformation $\xi_S$ earns its status as an ``edge-mode'' because it is  intrinsic to the boundary.\footnote{ There are different definitions of edge-modes in the literature, but being gauge-group-valued and intrinsic to the boundary is a common requirement.} But it is important to note that here it is not an independent degree of freedom at the boundary; it depends on the regional processes. 

Here we see Rovelli's handles at work: we may be given the physical transformations of the subsystems, $\hat{H}_{J}(\bb X_{J})$, but we still have the ``embedding'' information in the full field-space, which tells us, through  $\xi_S$, what sorts of ``rotations'' we need to perform at the boundary for the coupling, or ``language translation'', between subsystems. This ``translation'' is only possible because we are projecting onto the physical (or horizontal) subspaces but not discarding the ambient space and their plurality of perspectives.\footnote{ Note that essentially the same procedure would work at the intersection of charts, as in footnote \ref{ftnt:overlap}.}

There are many interesting questions we can ask about the properties of these projections viz a viz the composition of regions. For instance: what is the relation between the horizontal projections and the restrictions to the given regions? As is shown in \cite{GomesHopfRiello}, the two operations do not commute: given a globally horizontal  $\bb X=\hat H(\bb X)$, we may still find that its restriction is not regionally horizontal:
$$\bb X_1:=(\hat H(\bb X))_1\neq \hat{H}_1(\bb X_1)
$$
All this means  is that regionally one might still have to change frames so as to uncover the purely physical component of $\bb X_1$ relative to region and field-content. Thats is,  it means $\bb X_1$ has a non-trivial vertical component, which can be further removed.  In fact, it could even be that $\bb X_1$ is purely vertical:
$$\bb X_1:=(\hat H(\bb X))_1=\hat{V}_1(\bb X_1)\qquad \mbox{and therefore}\qquad \hat{H}_1(\bb X_1)=0$$
 in which case that perspective (i.e. regional configuration) \textit{experienced at most} a change of frames.\footnote{We don't mean `experienced'  in a conventional, physical sense. We mean solely that the region underwent a change of frames and nothing else.} However, it is always true that if the global $\bb X$ has a physical component (i.e. if $\hat{H}(\bb X)\neq 0$), then at least one region also has registered a  physical component to the transformation, e.g. $ \hat{H}_2(\bb X_2)\neq 0$. This fact lends objectivity to physical properties as measured by $\hat H$. 
 
Due to the possible lack of commutativity between restriction and projection, the global physical transformation does \textit{not} always decompose nicely into the local, purely physical transformations, i.e.  i.e. the following can fail to hold
 \be \hat{H}(\bb X)=\hat{H}_1(\bb X_1)+\hat{H}_2(\bb X_2)
\label{gluing} \ee
{ In fact, as shown in \cite{GomesHopfRiello} (see eq. 5.17 there), the necessary and sufficient condition for the validity of  the composition in \eqref{gluing} is precisely tracked by the discrepancy in rotations at the boundary, discussed in sections \ref{sec:Rovelli} and \ref{sec:contraRovelli}. Namely, \eqref{gluing} holds if and only if:
 \be \imath_S^*(\hat{V}_1(\bb X_1)):=\imath_S^*(\hat{V}_2(\bb X_2))\label{vertical_match}
 \ee
where $\imath_S^*$ is the pull-back to the boundary, acting on the spacetime indices of $A_\mu^I$ (or $\bb X_\mu^ I$) as usual, namely, for ${Y}^i\in \fX(S)$, $$(\imath_S^*\mathbf{A}^a)(\mathbf{Y})=A^a_\mu({\imath_S}_*(Y^i))^\mu=A^a_\mu \frac{\partial \imath_S^\mu}{\partial y^i}Y^i$$
where $y^i$ are intrinsic coordinates of $S$ and $x^\mu$ are those of $M$.  Therefore, for generic configurations, the physical decomposition \eqref{gluing} holds if and only if the RHS of \eqref{bdary_step} vanishes. That is: a physical (or horizontal) process precisely decomposes  into two physical processes if and only if the regional physical components match at the common boundary. Of course, for Yang-Mills theory this is by  no means a generic statement, it happens only for particular states (see also footnote \ref{ftnt:coincidence}). 

Taking \eqref{vertical_match} and \eqref{bdary_step} together, we have schematically shown that there exists an injective map:
\be\label{eq:injective}
H/(H_1\cup H_2)\rightarrow \text{Lie}(\G_S)
\ee 
I.e. the difference between global and local physical modes indeed has a correspondence to what one would call `edge-modes' (i.e. elements of $ \text{Lie}(\G_S)$). However, and this is important,  there is no implication  that our  ``edge-modes'' are independent degrees of freedom; each ``edge-mode''  will be completely determined by the regional processes. They are therefore of a distinct nature than the ones proposed in\cite{DonnellyFreidel}.

In sum, by  using $\varpi$,  the present framework keeps tracks of rotations of frames at the boundaries. It is therefore able to connect the two descriptions and diagnose the effects of a mismatch. 

	 \subsubsection{\textit{Quantum considerations} --}
{ For a classical system, it is indeed possible that the chosen perspectives give $\hat V_{1,2}(\bb X_{1,2})=0$ on each side, and therefore the horizontal components just match on the nose, as in choosing coordinates such that $\dot q^1_1=\dot q^2_N$ in Rovelli's example in section \ref{sec:Rovelli}, or such that \eqref{eq:gluing_simple} vanishes. However, once we pass on to the superpositions of the quantum domain,  we will have to account for $\bb X$ such that $\hat V_{1,2}(\bb X_{1,2})\neq 0$.}

When performing the path integral using a geometric decomposition (as in \cite{Mottola:1995sj}), integrations over regions sharing a boundary will not be completely independent. For instance, the second equation in \eqref{YM_varpi_boundaries} requires the projections of the vertical component normal to the boundary to be equal, and since the difference between horizontal projections is identified with minus the difference in vertical projections, as in \eqref{bdary_step}, these produce tractable correlations between the two regional path integrals. 
 My expectation (and of others that have worked on this topic) is that such intertwining between the two path integrals (one for each region) lead to the same results for entanglement entropy as was achieved in \cite{Agarwal},  without the use of additional edge-modes (or ``replica tricks''). 

{ In their role as  projectors of general degrees of freedom  onto the physical ones, the  connection-forms  closely resemble  BRST ghosts \cite{HenneauxTeitelboim}. There are two points of physical similarity: i) Feynman originally introduced ghosts from  cutting a process into parts and comparing it with the glued version (e.g. enforcing the Cutkosky rules for a non-Abelian theory) \cite{FeynmanGhost}. ii) In  the Kugo-Ojima quartet mechanism \cite{Kugo_Ojima}, the job of the ghosts is precisely to avoid counting as physical some degrees of freedom which are fundamentally gauge. Both of these points are familiar to us: the connection comes alive at boundaries, where it is used to identify and subtract the unphysical component of processes. Indeed, this work is motivated by  the confusion that boundaries engender in the splitting of physical and gauge degrees of freedom.   

At the mathematical level, the properties of BRST ghosts (its anticommuting nature and transformation properties) can be entirely reproduced by a flat $\varpi$---i.e. one associated to a gauge-fixing section. 
Moreover, according to geometrical interpretations of BRST \cite{Bonora1983, thierry1985classical, Thierry-MiegJMP}, ghosts can be associated to vertical one-forms and BRST transformations can be associated to vertical derivatives. If $c$ are the BRST ghosts, and $s$ the BRST transformations, the matching $s\leftrightarrow \dd_V$ and $c\leftrightarrow \varpi$ comes about as follows: 
$$\dd_V\varpi=-\frac12[\varpi,\varpi] \,\, \, (\text{BRST}: sc=-\frac12[c,c])\,\,\,\,\text{and}\,\,\,\, \dd_V A=\D_\varpi A\,\,\, (\text{BRST}:\, sA=\D_c A)$$
 In the integrable $\varpi$ case, both are `rigid' transformations which may be interpreted as  globally shifting the gauge-fixing section vertically (and therefore affording the required flexibility for matching of regions, see figure \ref{fig:distribution}).\footnote{ Having said that, since connection-forms come from taking derivatives in field-space, we do not currently know how to encode them in an action functional. For this reason, although we can fully reproduce the properties of BRST transformations ($\dd_V$) and BRST ghosts ($\varpi$), recovering the full \textit{BRST mechanism} is still vexing. \label{ftnt:BRST}} These discussions are included in \cite{GomesRiello2016}.}

\begin{figure}
		\begin{center}
			\includegraphics[scale=.4]{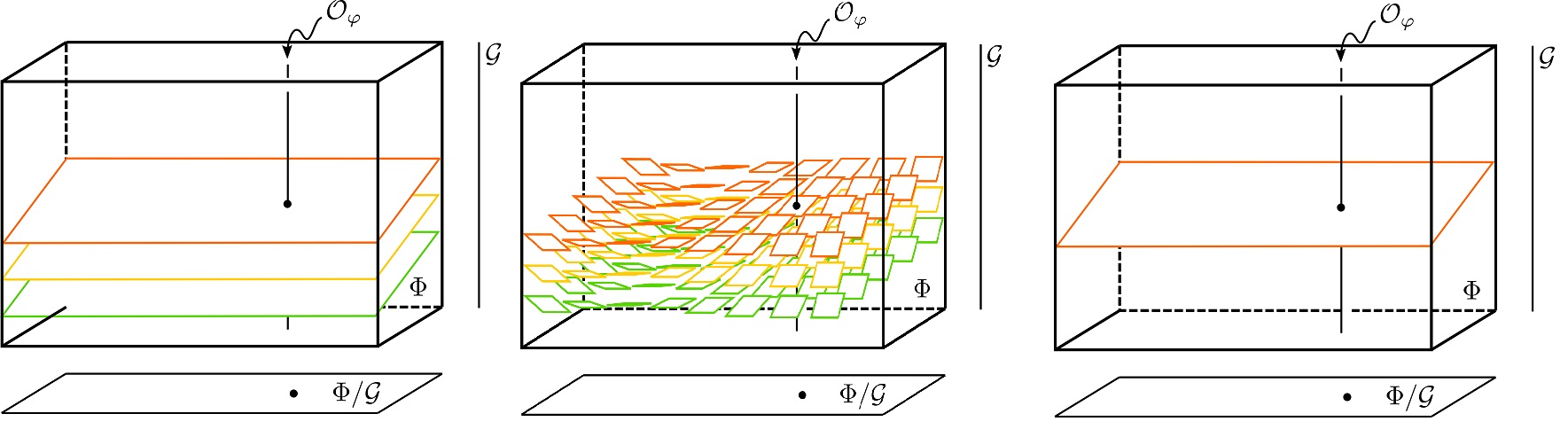}
			\caption{A visual illustration of the different structures provided respectively by: the rigid vertical  transformations compatible with BRST symmetry (but see footnote 42
			),  the horizontal distribution given by the connection-form, and  gauge-fixing.}
			\label{fig:distribution}
		\end{center}
	\end{figure}

\subsection{Horizontal Noether charges}\label{sec:hor_charges}
Having set up the stage, I will now illustrate the use of a connection-form explicitly in the computation of the Noether charges. We first focus on the general properties of the formalism, and then move on to a particular use of the SdW connection. 

		 \subsubsection{\textit{Symplectic geometry} --} 
Let $L = \mathscr{L}(\varphi)\d^d x$ be a  Lagrangian (spacetime-) {\it density}. We then define the presymplectic potential $\theta$ implicitly through the field-space derivative of this density: 
\be\label{eq:theta_implicit}
\boxed{\quad\phantom{\Big|}
\dd L = \mathrm{EL}_I(\varphi)\dd\varphi^I  + \d\theta(\varphi),
\quad}
\ee
where $\d$ is the spacetime exterior derivative, and $\mathrm{EL}_I(\varphi)$ are the (densitized) Euler-Lagrange equations for $\varphi^I$.

Let us for simplicity assume  the Lagrangian {\it density} and the symplectic potential are strictly invariant under (infinitesimal) gauge transformations, and not only up to boundary terms (at least for $\xi$'s which are field-independent)
\be
\fLie_{\xi^\#} L = 0,
\qquad\text{and}\qquad
\fLie_{\xi^\#} \theta = 0 \quad(\dd \xi = 0).
\label{hypothesis}
\ee
 Where $\bb L$ is the field-space Lie-derivative, and  along $\xi^\#$, i.e. along vertical directions.  The demands \eqref{hypothesis} are quite restrictive, but do apply to the standard Lagrangian and symplectic potential of Yang--Mills theory.

 Now, if \eqref{hypothesis} holds, then
\be
0 = \fLie_{{\lvf}^\#} L = \mathrm{EL}_I\delta_{\lvf}\varphi^I + \d\fI_{\xi^\#}\theta,
\label{eq_LieLagrange}
\ee
and one is led to define the Noether current density
$j_{\lvf}$ as (e.g. \cite{Iyer:1994ys}) 
\be
j_{\lvf} := \fI_{{\lvf}^\#} \theta \equiv \theta(\varphi,\delta_{\lvf}\varphi).
\label{eq_def_jxi}
\ee
Then, one has
\be
\d j_\xi \approx 0.
\ee
where $\approx$ assumes the equations of motion hold. i.e. the subspace of $\FYM$ defined by ${\rm EL}_I = 0$.
 
 Using the arbitrariness of $\xi\in\fG$, one concludes that the Noether current must be of the form
\be
\label{eq:j_is_dQ}
j_\xi = C_a\xi^a + \d Q_\xi \qquad \text{where}\qquad C_a \approx 0.
\ee
 This equation defines the charge density $Q_\xi$ whose relation to the total charge, $Q[\xi]$,  introduced in \eqref{superpotential_charge} is, in shorthand
 \be\label{eq:total_charge}\int j_\xi\equiv Q[\xi]\approx \oint Q_\xi
 \ee
  Equation \eqref{eq:j_is_dQ} is an instantiation of Noether's second theorem, which also shows the association between gauge symmetries and canonical constraints \cite{Lee:1990nz}, the latter being the canonical generator of the relevant gauge symmetries.  Explicitly, defining $\Omega:=\dd \theta$, we have  the Hamiltonian flow equation 
\be
\label{Ham_flow}
\fI_{{\lvf}^\#}\Omega = -\dd j_{\lvf} \qquad (\dd \xi =0).
\ee

Three remarks are in order: i) as can be seen in this case, the current $j_\xi$ is conserved automatically, but one cannot say much about the charge $Q_\xi$, which should be taken to be on a corner d-2 surface (usually taken to be a 2-sphere at infinity). Indeed, this is also an issue for Donnelly and Freidel's new charges---how can they be conserved in time?\footnote{One should devise new boundary action principles for them so that this is the case. Of course, one is then completely commited to the  material  particle content at the boundary.} ii) There is one charge per each choice of $\xi$. iii) We used $\dd \xi=0$ (gauge-transformations were taken to be field-independent) but, as we know, this can be quite restrictive. 

		 \subsubsection{\textit{Horizontal symplectic geometry} --} 

Why is considering $\dd \xi\neq 0$ at the boundary relevant? Boundary, or asymptotic conditions on the gauge-parameters  have an indelible reference to the physical field, e.g.: they are the ones that preserve certain properties of the fields; the two---gauge parameters and physical fields---are thus wedded at the boundaries. 

This can be confirmed in many ways.   For instance, when dealing with gauge-fixings, one must allow for $\dd \xi\neq 0$, e.g.: the orbit translation $\lambda^{\text f}(A)$ depends on the field. For instance, a one-parameter family of $\lambda(t)_{|\pp M}$ is equivalent to a one-parameter family of gauge-fixings, and  since a change of gauge-fixings is field-dependent (see e.g. equation \eqref{eq:gluing_simple}), one should also count a change of gauge at the boundary as being field-dependent (a change of gauge in the interior does not incur a change of gauge-fixing section, as seen in \eqref{gf_uni}). In other words, if one implements two different boundary conditions for the gauge-parameter at the boundary, these count as two ``possible worlds'', and therefore we can say without ambiguity that there exists a variation, $\delta \lambda_{|\partial M}\neq 0$ in the language employed in section \ref{sec:gf_bdary}, or $\dd\xi\neq 0$ at this more general, abstract stage.   


 While it is true that such a requirement---of  $\dd \xi\neq 0$---is no longer needed once the two bounded regions are glued and the boundary disappears, the same fate awaits the physical degrees of freedom taken to reside at boundaries (i.e. as $\G/\G^o$ for $\G_o$ given in \eqref{phys_edge}), which disappear once gluing is completed. A unified treatment---one that seamlessly treats bounded and unbounded regions---requires the consideration of $\dd \xi\neq 0$ at boundaries, and, for unification,  everywhere else too. 
 
 Just as dealing with spacetime dependent gauge transformation required the  original finite-dimensional connection, dealing with field-dependent gauge transformation requires the field-space connection-form. Appropriately dealing with $\dd\xi\neq0$ suffices for finding a precise definition of horizontal symplectic geometry \cite{GomesRiello2016}.  

What happens when  (iii) is dropped? Many of the previous computations fall apart. For instance, we no longer have the relation \eqref{Ham_flow}. The obstructions can be traced back to a violation of the conditions \eqref{hypothesis} for $\dd \xi\neq 0$, 
\be
\fLie_{\xi^\#} \theta = \fI_{\xi^\#}\dd \theta + \dd \fI_{\xi^\#}\theta \;\stackrel{\eqref{hypothesis}}{=}\; \fI_{(\dd\xi)^\#}\theta = j_{\dd \xi} \approx \d Q_{\dd \xi}.
\ee
It was by eliminating this obstruction that Donnelly and Freidel got rid of the arbitrary charges related to the local gauge symmetries (only to replace them with another infinite set of  charges). 

In our case, following a more geometrical approach to covariant symplectic geometry leads us to the introduction of the horizontal symplectic current  \cite{GomesRiello2016}:
\be
\boxed{\quad\phantom{\Big|}
\theta_H :=\Pi_I\dd_H\varphi^I = \theta - \fI_{\varpi^\#}\theta = \theta - j_\varpi.
\quad}
\label{defhorizontalpotential}
\ee
If $\theta$ satisfies the conditions of equation \eqref{hypothesis}, then---even for a field-dependent $\xi$ ($\dd \xi \neq 0$)---one has automatically $\fLie_{\xi^\#}\theta_H  = 0$. I.e. the charges will be completely gauge-invariant. 

Thus the horizontal presymplectic two-form is automatically $\dd$-exact, and differs from the standard symplectic form by a spacetime exact term ($\alpha$ is a field-space one-form and spacetime scalar):
\be
{\quad\phantom{\Big|}
\Omega_H:=\dd_H\theta_H=\dd\theta_H=\Omega+\d\dd\alpha,
\quad}
\label{horizontalOmega}
\ee
In other words,  $\Omega_H$ is $\dd$-closed and therefore a viable presymplectic form.

From the above
\be
\phantom{\Big|}
j^H_{\lvf} := \fI_{{\lvf}^\#}\theta_H =  \fI_{{\lvf}^\#}\theta - j_{ \fI_{{\lvf}^\#}\varpi}=j_\xi-j_\xi=  0
\label{hor_charge_gauge}
\ee
where we employed \eqref{varpi_compl}, and horizontality (but no gauge-fixing) was used to annihilate the local gauge charges. 

It is  similarly easy  to see that $\fI_{{\lvf}^\#}\Omega_H = 0$ from $\fLie_{\xi^\#}\theta_H  = 0$ and the Cartan magic formula. These formulas are valid at the density level, and hold for field-dependent gauge transformations as well.
The message they convey is that such gauge transformations carry {\it no physical charge} with respect to $\varpi$. 

Indeed, local gauge transformations, unlike global gauge transformations, should not produce meaningful physical charges. Whereas the standard symplectic potential does not make this distinction---naively assigning non-trivial currents to both types of transformation through Noether's theorems---here we seem to have the opposiste problem: our modified symplectic potential apparently leaves no charge at all!

\subsubsection{\textit{Global charges} --}

For  transformations  which respect \eqref{eq_fundamental}, $\varpi$ correctly identifies the pure gauge ones and  \eqref{hor_charge_gauge} eliminates the respective charge. This procedure is encapsulated by the validity of  $j^ H_\xi=0$ for such $\xi$. And \eqref{hor_charge_gauge} holds for all bona-fide connection-forms in principal fiber bundles. However,  $\F$ is not in general a bona-fide principal fiber bundle, and the relational connection-form $\varpi$ is not in general a bona-fide connection-form. 

Let us take a look again at \eqref{hor_charge_gauge}. It is easy to see that if there existed $\xi_o\neq 0\in \fG$ such that $(\varpi(\xi_o^\#))^\#=0$ (and therefore  \eqref{varpi_compl} did not hold completely), then we could conceivably be left with a non-trivial charge for some such $\xi_o$.\footnote{\label{ftnt:GW} In \cite{GreavesWallace}, a similar criterion is proposed for the definition of `empirically significant symmetries'. In the language employed here (in \ref{sec:gluing} in particular),  the conditions there would be formulated, in infinitesimal form, as: a transformation $\bb X$ is an empirically significant symmetry if  for two regions, $M_1\cup M_2=M$,  $\hat H(\bb X)\neq 0$ but $\hat H_1(\bb X_1)=\hat H_2(\bb X_2)=0$. The necessary property they find for this is that there exists a $\xi$ such that $\xi^\#_{|\pp M}=0$.  An analysis of this framework in our language shows that no such empirically significant symmetry exists in finite bounded regions corresponding to local gauge symmetries \cite{GomesRiello_des}.} 
\textit{If $\xi_o$ is in the blind-spot of $\varpi$, and the standard charge for it does not vanish on-shell, $j_{\xi_o}\not\approx 0$, then}
\be j^H_{\xi_o}=j_{\xi_o}\not\approx 0\ee

\subsubsection{\textit{The point of confusion} --}

 Usually, there is no essential difference between changes of frame and changes of perspective, i.e. between $\xi$ and $\xi^\#$ ---there is a bijection between $V_\varphi$ and $\fG$. One essential point which has confused discussions in both the philosophy and physics communities is that this bijection does not always hold: the group of \textit{effective} gauge transformations can differ in important ways from the group of gauge transformations.
 
 And indeed,  if global gauge transformations are merely specific cases of the local gauge-transformations, how are they to escape extermination by  \eqref{hor_charge_gauge}? The answer: by exploring the aforementioned blind-spot of perspectives. 

  The global group is defined from the local one---usually in a field-dependent manner---through $\varpi$  by a projector onto the kernel of a differential operator.  
 For instance, in  electromagnetism taking $\xi={const}$ we have  $\xi\neq 0, \varpi(\xi)^\#=0$, for all base points $A_\mu$. There, the projection is field-independent, but this is not the case in general, since in general the group action is itself field-dependent (e.g. $\xi^\#=\D \xi$ for Yang-Mills, with $\D$ containing $A_\mu^I$). For such group-actions, finding all the `blind-spots' becomes less trivial. Nonetheless, we can safely say that blind-spots always contain \textit{the stabilizer subgroups} of given configurations, i.e. elements $g\in \G$ such that $R_g\varphi=\varphi$. 
But it is not every $\varphi$ which has a non-trivial stabilizer. Much to the contrary: In the space of non-Abelian potentials and gravitational fields, generic configurations have no stabilizer subgroups. 

Nonetheless, stabilizers do exist. Therefore  the orbits $\mathcal{O}_{\varphi}\subset \F$ are not isomorphic for \textit{all} $\varphi\in \F$. Indeed, different stabilizer groups produce qualitatively different orbits. When taking the quotient to obtain the space $\F/\G$, such discrepancies show up in the topological structure itself: $\F/\G$ is not  a manifold, but only a stratified manifold, a nested union of submanifolds of different dimensions \cite{Fischer, kondracki1983}.

Global charges appear from these irregularities in $\F/\G$; they present themselves as transitions between different regions. More specifically, global charges appear when $\F_\varphi/\G$ for a  given field $\varphi$ has such a transition, and another sector does not (we need the standard $j_\xi\not\approx 0$, as we see below, in equation \eqref{Noe_charge}). 

In essence, the connection-form is  finding a projection from the local to the global transformations, and the modified symplectic charge gives us the corresponding physical charge.

\subsubsection{\textit{`Slippage': an illustrative example} --}
Again, in certain respects examples with non-internal symmetries are more palpable than the ones with internal symmetries. Therefore it useful to get the gist of what is to come through a heuristic example of how horizontal charges arise from the qualitatively different actions of the gauge group on different physical fields. I.e. they arise from the relational non-triviality of `blind-spot transformations'. 

For example, generic spacetime metrics have no non-trivial Killing vector fields, and therefore have no non-trivial conserved charges. 
For the few orbits that do possess isometries, the group of diffeomorphisms acts in a qualitatively different manner. Namely,  there is some `slippage' of the action of the group on the orbit: a finite-dimensional subgroup $\mathcal{D}_o\subset \text{Diff}(M)$ does not move some metrics--- those which are said to have stabilizer group $\mathcal{D}_o$, and which we can call 
\be I_o:=\{g^o_{ab}\,|\, f^*g^o_{ab}=g^o_{ab}\,\, \text{for}\,\, f=h\circ f' \circ h^{-1},\,\, h\in  \text{Diff}(M), f'\in  \mathcal{D}_o\}.\ee
 For example, for Minkowski metrics, $g^o_{ab}=\eta_{ab}$, in which case $\mathcal{D}_o=\mathcal{P}$ is the Poincar\'e group.\footnote{Note that in the definition of $I_o$ must be independent of the representation of the metric, i.e. a metric $h^*\mathbf{\eta}$ will have isometries given by $h\circ f' \circ h^{-1}$ where $f'$ are the isometries of $\eta$.}  Reciprocally, note that for the Poincar\'e group to be well-defined in this manner one must restrict attention to elements of the corresponding $I_o$.\footnote{This is why I disagree with Greaves and Wallace, when they say that ``In any theory that has a \lq{}local\rq{} symmetry group, the \lq{}global\rq{} symmetries
remain as a subgroup of that local symmetry group. (For example, in general
relativity, the \lq{}global\rq{} translations and boosts form a subgroup of the group of all
diffeomorphisms.)'' \cite{GreavesWallace}. It is only for highly homogeneous metrics that this subgroup is unambiguously defined.}
 
 Suppose now that we don't have just Minkowski spacetime, but also a particle within it. Poincar\'e symmetries leave the Minkowski metric completely invariant but have a non-trivial action on the particle. Therefore, apart from Poincar\'e transformations---which should produce charges by the mechanism above---generic diffeomorphisms should not produce any charge.\footnote{A caveat: we have not yet completed the work on relational symplectic geometry for gravity. These statements are based on an assumption that the gravity case will work as well as the Yang-Mills one. }

\subsubsection{\textit{Conserved global charges: an explicit example} --}

Finally, we can see in an explicit example how the horizontal symplectic geometry due to the SdW connection-form filters out the global charges from the local gauge transformations.

We start with the Yang-Mills Lagrangian in form language:
\be
L_\YM = -\tfrac{1}{2e^2} (F^a \wedge  \ast F_a) +  (i \bar\psi\gamma_\mu \D_\mu\psi) \wedge \ast \d x^\mu 
\ee
where $F = \d A + \tfrac12[A,A]$ and $\ast$ is the Hodge dual, $\gamma^\mu$ are Dirac's gamma-matrices, $\D_\mu \psi = \pp_\mu \psi + A_\mu^a \tau_a \psi$,  in the $\tau_a$ algebra  basis, and $e$ is the Yang--Mills coupling constant. For simplicity of notation, we have also defined $\bar\psi := \psi^\dagger \gamma^0$.

One can find the Noether current by the standard method, as: 
\be j_\xi:=\fI_{\xi^\#} \theta_\YM = - e^{-2}\d (\ast F_a \xi^a)=\d(\text{Tr}(E{\xi})) \label{Noe_charge}
\ee
where we identified $\ast F_a$ pulled-back to the d-1 surface  with the electric field (and also used index-free notation and absorbed the coupling constant), and the standard symplectic potential is
\be\label{theta_YM}
\theta_\YM =  -e^{-2}\dd A^a \wedge  \ast F_a + i \bar\psi \gamma_\mu \dd \psi \ast \d x^\mu.
\ee
Although $\theta_\YM$ in \eqref{theta_YM} is invariant under field-independent gauge transformations (as demanded in the previous section), under field-dependent ones: 
\be{\quad\phantom{\Big|}
\fLie_{\xi^\#} \theta_\YM = -e^{-2}\d (\ast F_a \dd\xi^a )=\d(\text{Tr}(E{\dd\xi}))\not\equiv 0.
\quad}\label{Lie_theta_YM}
\ee

Skipping technical details (contained in \cite{GomesHopfRiello}), let $\varpi$ be associated to one field, the reference-field, say the gauge potential $A$, for which we choose the Singer-DeWitt connection \eqref{eq:sdw}.

Now the non-zero term in \eqref{Lie_theta_YM} can be eliminated by using the horizontal (or physical wrt $\varpi$) symplectic potential, $\fLie_{\xi^\#}\theta_{{\rm YM}, H}=0$, where 
\begin{align}
\theta_{{\rm YM}, H}& = - e^{-2}\dd_H A^a \wedge \ast F_a + i \bar\psi \gamma_\mu \dd_H \psi \ast \d x^\mu= \theta_\YM -\d(\text{Tr}(E{\varpi}))\\
\Omega_{{\rm YM}, H}&=\dd_H\theta_{{\rm YM}, H}=\dd\theta_{{\rm YM}, H}=\Omega_{\rm YM}+\dd d(\text{Tr}(E{\varpi}))
\end{align}
Note that the modification from the Yang-Mills symplectic form is both spacetime and field-space closed (i.e.  $\d \dd \alpha$), which is a guarantee that it defines a (pre)symplectic geometry and that we recover the standard Yang-Mills symplectic geometry in the absence of boundaries. Moreover, \textit{generically}   $\fI_{\xi^\#}\theta_{{\rm YM}, H}=j^H_\xi=0$.  

So far in this example, we have obtained all the desirable results of introducing an `edge-mode', in the spirit of Donnelly and Freidel \cite{DonnellyFreidel}. Their symplectic potential is given by
\be \theta^{\text{DF}}=\text{Tr}(E\delta A)-\d (\text{Tr}(E\delta \kappa \kappa^{-1}))
\ee
whose shape we  recover with the Higgs connection, $\varpi = - \dd h   h^{-1}$, given in \eqref{varpim_flat}. What are the differences?

 In their nomenclature, Donnelly and Freidel recover charges from a different transformation of their edge-modes, which they call ``symmetry transformations'' \cite{DonnellyFreidel}. In general, these are not available to us.\footnote{Essentially gauge transformations are given by a left action and symmetry transformations by a right one, as seen in \eqref{surface_symmetries} and \eqref{surface_gauge}. The meaning of these transformations on the new degrees of freedom inhabitating the boundary are unclear to me. If they are somehow supposed to correspond to  choices of `observers', why constrain such transformations to be related to the gauge group? }
 As mentioned before, they recover a plethora of charges at any given boundary, related to the intrinsic boundary degrees of freedom, and I consider this to be an unphysical overshoot; it depends on much more than the original fields available to us. But how do \textit{we}  recover charges using the formalism here developed?

We note that the matter field $\psi$  does not share all its stabilizer subgroups with the reference field. I.e. the field whose perspective we are using to map changes  has blind-spots which are not shared by another charged field. 
In our case, as shown in \cite{GomesHopfRiello} and described in the previous section, $j^H_\xi=0$ for $\xi$ which are \textit{not} stabilizers of $A^a$. However, if there exists a set of Lie algebra elements $\{\chi^a_n\}$ forming a basis for the stabilizer of $A^a$, such a transformation will  not annihilate a non-vanishing matter field, since $\delta_{\chi_n}\psi=-\chi_n\psi\neq 0$. From this and the above remarks, for such a ``Killing''  transformation $\chi_n$ we obtain
\begin{align}
j^H_\chi =  -\chi^a i \bar\psi \gamma_\mu \tau_a * \d x^\mu =  \chi^a J_a
\qquad
(\D \chi = 0).
\end{align}
where we introduced the matter current density: \be J_a = - i  \bar\psi \gamma_\mu \tau_a\psi (\ast \d x^\mu) \ee

Thus, the horizontal current for a global transformation is precisely given by the {\it matter current density $J_a$} contracted with the globally defined $\chi^a_n$. In electrodynamics, $\chi = \text{const.}$ and $j_\chi^H$ is precisely the total current density of electrons.

Using the Gauss law, $D \ast F_a + J_a \approx 0$ {\it and} the Killing condition $\D \chi_n =0$, the horizontal charge can  be written as
\be
Q^H[{\chi_n}] =\int_\Sigma \chi_n^a J_a \approx -e^{-2}\int_{\pp \Sigma} \ast F_a \chi^a_n
\qquad
(\D \chi = 0).
\label{eq:horiz_charge}\ee
 We thus see that {\it the SdW connection with boundary picks out the global charges---when they exist---as the only physical ones}.

While the standard $j_\xi$ is non-trivial for any gauge transformation, $j^H_\xi$ is trivial for all, except for a few `global' gauge transformations. Ultimately, horizontal charges appear related to objectively conserved physical quantities, like the total charge in electromagnetism, while the standard Noether charge $Q[\xi]$ can be non-trivial for any gauge parameter.

 The space generated by all the blind-spot directions of a given perspective will have a maximum, finite dimension (like the 10-dimensional group of Poincar\'e transformations for the metric, or the one-dimensional group of constant potential shifts in electromagnetism). This procedure therefore provides a projection from dizzying infinite-dimensionality of local gauge transformations, to a finite number of gauge transformations---the `global ones'. For these, we cannot vary the gauge parameters point by point, and  Noether's first theorem applies in  standard fashion. Importantly, when obtained from the connection-form, these charges do not magically appear once one conjures a boundary, but correspond to physical degrees of freedom living in both bulk and boundary. They emerge from  symmetries which act qualitatively differently on each of the  interacting fields. 

\section{Conclusions}

I will now briefly summarize our trajectory through  this paper:
\subsection{Summary}
\subsubsection{\textit{Perspectives} --}
 We started by relating the glitz of fiber bundles to a more ``down-to-earth'' understanding related to perspectives---essentially arbitrary frames in which we describe our system. 

Because we have no local description of the physical degrees of freedom of gauge theories,\footnote{E.g. we could try to use holonomies as basic variables, but these are  non-local and cumbersome to couple to sources \cite{Healey_book}.} we use particular (but arbitrary) perspectives. Nonetheless, such a plethora of perspectives is usually seen as superfluous, since the assumption is that we can always gauge-fix to a single class and thereby attain a gauge-invariant description. 

In a field-theoretic language, one always assumes access to the gauge-invariant space, $\F/\G$. But this disregards regionality: one may no longer be able to couple regional  reduced, quotient physical spaces---once you have quotiented, your description becomes irretrievably rigid. 

In other words, when our access to the system is only regional, not global,  the survival of some perspectival flexibility is compulsory for describing the coupling of subsystems. It seems gauge degrees of freedom really come into their own when, as Rovelli advocates, they are related to subsystems \cite{RovelliGauge2013}. This argument suggests that describing  subsystems in particular gauge-fixings may be premature if we allow for limited access to the Universe.  We need to treat field-theoretic systems in the entirety of the field-space framework, and then extract whatever physical information we can from that larger domain.

\subsubsection{\textit{Noether charges} --}
We can track the physical effects of this extra information by their impact on the computation of conserved charges---the teeth of symmetries found by Emily Noether. When using symplectic methods in finite regions, this extra information at the boundaries can  correspond to an infinite number of  charges. Such charges depend on unphysical choices---i.e. not associated to perspectives---of gauge generators at the boundaries. Therefore, the physical status of these charges is dubious at best, unlike what is the case for those corresponding to  rigid, global symmetries.

It is not easy to go from these Noether charges of local gauge symmetries to ones generally recognized as physically meaningful conserved charges. 
An attempt by Donnelly and Freidel leaves the boundary information completely free, by introducing `edge-modes' \cite{DonnellyFreidel}. These degrees of freedom eat up the spurious conserved currents associated to the local gauge symmetries, but they generate another infinity of conserved currents which are now dressed by new degrees of freedom---i.e. not in the original field-content---existing only at boundaries (imagined or not). 
 Having the correct covariance properties for coupling regions, they could be seen as an implementation of Rovelli's handles. However, these handles are being attached to the system as extra baggage. 

\subsubsection{\textit{The relational connection-form} --}
The connection-form, $\varpi$, follows Rovelli's idea, finds something similar to Donnelly and Freidel's construction, and ends up with physically meaningful notions of charges. For $\varpi$ is a device that fills three needs: i) providing a reference to measure \textit{the physical  change} of states, $\hat H(\bb X)$; ii) keeping track of the perspective from which it does this in a covariant manner (allowing for the gluing of regions), $\hat V(\bb X)$; and iii) finding relational physical charges  requiring solely the physical fields, $Q^H[\chi_n]$ in \eqref{eq:horiz_charge}. Let us quickly review these points.

The connection-form does not provide a fixed  selective class of perspectives  for comparing arbitrary states---as a gauge-fixing would,---it only provides a reference for examining infinitesimal changes in a given state (see figure \ref{fig:distribution}, for an illustration in terms of non-integrable distributions in $\F$).  $\varpi$ keeps track of the change in perspective by  telling us how the fields extend an arbitrary choice of frame infinitesimally into the future. It then pinpoints any unwitting rotation of our arbitrary frame in the description of the process.  Therefore it implicitly also tells us how much our  system \textit{physically} changed (infinitesimally) with respect to the given  field and perspective, and according to some norm on field space (e.g. the supermetric appearing in the kinetic term in the Lagrangian).  One of its key properties is that it is able both to select notions of physical change \textit{and} negotiate the gluing of regions. 

 $\varpi$  is also exclusively dependent on the field-sector and on the region of space occupied by the physical system; no extra degrees of freedom are required. For this reason I call it a ``relational connection-form''. 

Mathematically, $\varpi$ defines a notion of horizontality in field-space. A horizontal change in the field corresponds to a change that is `physical with respect to the perspective'.  Although it is not strictly relational in the sense of Rovelli---i.e. ``requiring another subsystem to be made sense of''--- $\varpi$ also comes into its own in the presence of boundaries coupling to different regions.

 With a connection-form, we are also able to define an alternative (pre)symplectic structure in the space of fields. The horizontal symplectic geometry defined with the aid of $\varpi$ only differs from the standard symplectic geometry in the presence of boundaries to the system.\footnote{By boundaries here I mean $d-2$ surfaces, boundaries to the spatial hypersurface, sometimes called `corners'. } 
 Nonetheless, this boundary difference in symplectic structure is enough to annihilate charges associated to local gauge symmetries, irrespective of boundary conditions. Fortunately, the formalism rescues the baby before throwing out the bathwater: it filters out  charges associated to global symmetries.  These charges are associated to particular irregularities in the `physical base space', $\F/\G$, and are due to the second instance of detachment between  gauge symmetries and perspectives.

 Usually, not much attention is paid to the structure of the reduced space $\F/\G$. It is assumed to abstractly represent the view-from-nowhere of the physical content of the system, without any redundancy. But  $\F/\G$'s irregularities are physically relevant, representing the blind-spots of perspectives, and without first describing the full configuration space  it is impossible to find them; after all, blind-spots emerge from transformations of perspective which are not registered by the field in question. In this respect, the horizontal charges resemble the definition of ``empirically significant symmetries''  of \cite{GreavesWallace} (see footnote \ref{ftnt:GW}), but one more relational fact is needed to make conditions ripe for global charges: a given subsystem's `blind-spots' must not be shared by the coupled subsystem. The emerging charges are therefore of a patently relational character; they crop up when one subsytem slips from the other during a gauge transformation. A trivial example would be constant gauge transformations in electromagnetism in the presence of  charged matter.
 
 \subsection{Open questions} There are many questions posed by boundaries in gauge theory; no serious thinker about gauge theories should skirt these questions. The formalism described in the second half of this paper is still young; it may successfully deal with the sampling of questions described in the first half of the paper, but multiple questions remain. Here I will only briefly comment on what are for me the most pressing ones. 
 
 The inherent tension between locality and gauge invariance is usually  expressed by the failure to factorize the physical Hilbert space into local tensor products. This failure is relevant in studies of entanglement entropy \cite{Donnelly_entanglement}. In fact, one of the main motivations for \cite{DonnellyFreidel} were computations of entanglement entropy in gauge theories \cite{Donnelly_entanglement}: when one is tracing over all degrees of freedom on one of the regions;  extra information at the boundary---such as e.g.: fixing $\lambda$ by hand---need to be counted. As mentioned in footnote \ref{ftnt:Agarwal}, the expectation is that we can recover the correct computations of entanglement entropy through path integral methods (without the use of replica tricks, as is done in \cite{Agarwal}), but this remains to be seen. 
 
 Another glaring gap in our approach is the adaption of the treatment to the diffeomorphisms of general relativity. This is work under way and I see no reason to doubt its success. 
 
 A third important open question is the adaptation of the formalism to restrictions on field-space, as occurs when stringent boundary conditions are imposed on field-content---a very useful tool to study particular regimes or types of subsystems. Our work shows that boundaries by themselves may generically have no charges associated to them (e.g. for non-Abelian theories).  But it does not mean that a) there are no charges for a particular field-content. The charge comes from the field, not from the boundary. Particular field-contents which have charges associated to them include i) base-point field-configurations with stabilizers but no restriction on field-variations, and, hopefully, ii) restrictions on field-space (e.g. boundary conditions). The open problem is to show that (ii) really does come to pass once one considers the appropriate restrictions in the derivation of $\varpi$.  
 
 Finally, we have mostly used a single manifold chart to describe our constructions. This assumption is also not compatible with regional access to the World. The work done in \cite{GomesRiello2016, GomesRiello2018, GomesHopfRiello} needs to be extended to more general topologies of the underlying manifold. 


\subsection*{Acknowledgments}

I would like to thank  Florian Hopfmuller, and especially Aldo Riello,  co-authors  of the papers  on which this analysis is almost entirely based. Any mistakes of interpretations not found in the original papers are of course my own. I would also like to show my immense gratitude to Jeremy Butterfield, for a meticulous reading, offering invaluable comments and advices on a first draft. 

 \bibliographystyle{alpha}

\end{document}